\documentclass[letterpaper,aps,prb,floatfix,twocolumn,showpacs,superscriptaddress,10pt]{revtex4}
\pdfoutput=1
\usepackage{graphicx}

\usepackage{amssymb,amsmath}
\usepackage{multirow}
\usepackage{color}
\usepackage{bm}
\usepackage{longtable}
\usepackage{lipsum}
\usepackage{makecell}
\usepackage{array, multirow, hhline}

\def\be{\begin{equation}}
\def\ee{\end{equation}}
\def\ba{\begin{eqnarray}}
\def\ea{\end{eqnarray}}

\begin{document}

\title{Perfect quantum excitation energy transport via single edge perturbation in a complete network}

\author{Hassan Bassereh}
\affiliation{Department of Physics, Isfahan University of Technology, Isfahan 84156-83111, Iran}

\author{Vahid Salari}
\affiliation{Department of Physics, Isfahan University of Technology, Isfahan 84156-83111, Iran}
\affiliation{Foundations of Physics Group, School of Physics, Institute for Research in Fundamental Sciences (IPM), Tehran 19395-5531, Iran}

\author{Farhad Shahbazi}
\affiliation{Department of Physics, Isfahan University of Technology, Isfahan 84156-83111, Iran}

\author{Tapio Ala-Nissil{a}}
\affiliation{Department of Applied Physics and COMP CoE, Aalto University School of Science, P.O. Box 11000,
FI-00076 Aalto, Espoo, Finland}

\date{November 4, 2015}

\begin{abstract}
We consider quantum excitation energy transport (EET) in a network of two-state nodes in the 
Markovian approximation by employing the Lindblad formulation. We find that EET from an initial 
site, where the excitation is inserted to the sink, is generally inefficient due to the inhibition of transport 
by localization of the excitation wave packet in a symmetric, fully-connected network. 
We demonstrate that the EET efficiency can be significantly increased up to $\approx100\%$ by perturbing hopping transport between the initial node
and the one connected directly to the sink, while the rate of energy transport is highest at a finite value of the hopping parameter. 
We also show that prohibiting hopping between the other nodes which are not directly linked to the sink does not improve the efficiency. 
We show that external dephasing noise in the network plays a constructive role for EET in the presence of 
localization in the network, while in the absence of localization
it reduces the efficiency of EET. We also consider the influence of off-diagonal disorder in the hopping parameters
of the network.
\end{abstract}

\pacs{03.65.Yz,	
03.67.-a 
}

\maketitle
\section{Introduction}
\label{introduction}
There are many systems in nature and society that can be modelled by classical complex networks and explained by
statistical physics.\cite{Junker.2008, Rubi.2003} Recently, the quantum world has opened up new perspectives in the field of complex networks.~\cite{Biamonte.SCi2013, Caruso.New2013, 12} 
For example, energy, charge, or information transfer are important phenomena in physical and biological 
systems taking place at scales ranging from atoms to large macro-molecular structures, and the idea has been put forward that 
quantum mechanics might have a positive effect on the efficiency of energy or charge transport in such systems. 
Charge transport through DNA~\cite{1} and energy transfer in photosynthetic structures~\cite{2, 3, fleming.Science2007, 5} 
are good examples in this context. In fact, the most important effect of quantum mechanics in biological systems to date 
has been seen in the Fenna-Matthews-Olson (FMO) complexes~\cite{6},  observed by experimentally via 
ultrafast spectroscopy~\cite{7}, where there is an ultrahigh efficient excitation energy transport (EET) in 
light-harvesting complexes.\cite{fleming.Science2007, scholes2010,7, hulst.Science2013, Johnson.PRB2008, plenio.The Journal of Chemical Physics2009, Plenio.NJPH2010} This can be modeled by quantum walks, \cite{Mohseni2008, Mohseni2009, Zagury.PRA1993, Blumen.Phys2011} or the Lindblad formalism.\cite{plenio.The Journal of Chemical Physics2009, Plenio.NJPH2010} Simulated artificial complex networks based on biological systems have the potential to be used in future quantum informational and computational technologies such as secure information transfer between two particles in an entangled quantum-cryptography,\cite{Brassard.ieee1984} teleportation in a quantum communication protocol,\cite{Bennett.PRL1993,Lewenstein.Nature2007,Wehr.PRA2008} artificial photosynthesis systems (or solar cells) to save energy, and quantum neural networks\cite{Kimble.Nature2008,Thew.Natyrphoton2007} for advanced technologies based on artificial intelligence.

The influence of edge deletion in networks for quantum state~\cite{tsomokos, bose} and 
energy transfer~\cite{novo} has been studied; however, the influence of edge perturbation and deletion on 
the destructive interference of transition amplitudes in a network \cite{plenio.The Journal of Chemical Physics2009}
remains unexplored. In this paper we consider this problem 
in a fully connected network and show how the energy can be transferred in the network via deletion of a single edge.
We analyze quantum energy transport in a simple model of a complete network of two-state nodes. In particular, we demonstrate how energy transfer in such a network, which is inhibited by localization due to destructive interference, can be enhanced by breaking the symmetry between the nodes, or introducing disorder in the
coupling between the network nodes. Our work is relevant from the point of view of building an artificial network for high efficient energy transfer to simulate systems such as FMO complexes in photosynthetic structures. 


The paper is organized as follows. In Sec.~\ref{eet}, we describe the model Hamiltonian of the network and introduce the method of calculating excitation energy transport, employing the Lindblad formulation. In Sec.~\ref{fcn} we investigate EET in a 
fully connected network and discuss the negative effect of localization in this case. In Sec.~\ref{edge-deletion} we study the influence of 
single edge deletions in the network on the efficiency of EET. We then investigate the influence of dephasing on EET in Sec.~\ref{dephasing}, 
by considering edge-deletions. In Sec. \ref{saturation} we examine how the rate of energy transport can be optimized in the network. Finally, in Sec. \ref{Off-DiagonalDisorder} we discuss the role of off-diagonal disorder
in the coupling constants between the network nodes in enhancing the efficiency. Section
\ref{conclusion} presents our summary and conclusions.

\section{Quantum excitation energy transport in networks}
\label{eet}

Consider a graph $G$ (i.e. a complex network) as a pair of sets $G=(\nu,\varepsilon)$, where $\nu$ is a set of \emph{vertices} or \emph{nodes} of the graph and $\varepsilon$  is a set of \emph{edges} or \emph{links}  $V_{ij}$ ($V_{ij}\in \varepsilon$) connecting  the vertices $i$ and $j$ ($i,j \in \nu$).  An undirected  graph is completely defined by its
adjacency matrix $A$, defined as:
\begin{equation}
A_{ij} = \left\{ {\begin{array}{*{20}c}
   {0,\:\:\:\:\:\:\:\:\:  \text{for}\:\:\:\:\: V_{ij}\notin \varepsilon ;   } \\
   {1,\:\:\:\:\:\:\:\:\:  \text{for}\:\:\:\:\: V_{ij}\in \varepsilon . }  \\
\end{array}} \right.
\end{equation}
A \emph{complete $N$-graph} (i.e. a fully connected network) is a graph with $\binom{N}{2}$ edges where $V_{ij}=1$ for the all pairs of nodes\cite{BARRAT.CUP2008}.

Consider a network, consisting of $N$ nodes, in which each node is a two-state object such as a molecule or a qubit, with a ground and an excited state. We assume that the nodes 
interact with each other through direct hopping. When hopping between two molecules is allowed, a link is drawn between them.    
Whenever an excitation is inserted at one node, it can then be transfered throughout the network by hopping due to the interaction between the linked nodes. 
Here we consider quantum energy transport in a network that can be modelled by the following tight-binding Hamiltonian~\cite{plenio.New.J.Phys.2008}:
\begin{equation}
\label{TBH}
H=\sum_{n=1}^{N} \hbar \omega_n |n\rangle\langle n|+\sum_{n\neq m}^{N}J_{nm}(|m\rangle\langle n|+|n\rangle\langle m|) ,
\end{equation}
where $|n\rangle$ is the $n$-th site in which the excitation exists, $\hbar\omega_{n}$ denotes the excitation energy at site  $n$,  and $J_{nm}$ is the hopping integral between the two 
sites $n$ and $m$.  We set $J_{nm}=1$ when the nodes $n$ and $m$ are connected and $J_{nm}=0$ if they are disconnected, and we choose $\hbar=1$ so that
all the energies are in units of $\omega_{n}$.
Possible decay events of the exciton to the ground state are neglected here.

To study dissipationless quantum excitation energy transport (EET) in the network, we use the evolution of the master equation in the Markovian approximation as follows \cite{hassan,ali}:
\begin{equation}
  \dot{\rho}=-i[H,\rho]+L_{\rm{sink}}\rho,
  \label{master}
\end{equation} 
 in which $\rho $ is the density matrix.
$L_{\rm{sink}}\rho$ is the \emph{sink} term that expresses the irreversible transfer of energy from a given node of the network into a sink, and it is defined as:
\begin{equation}
\label{sink}
L_{\rm{sink}}\rho=\Gamma [2{\sigma}^+_{\rm{sink}} {\sigma}^-_{\rm{f}}\rho{{\sigma}^+_{\rm{f}}} {\sigma}^-_{\rm{sink}} -\big\{{{\sigma}^+_{\rm{f}}{\sigma}^-_{\rm{sink}}{\sigma}^+_{\rm{sink}}{\sigma}^-_{\rm{f}},\rho}\big\}],
\end{equation}
where the curly brackets denote the anticommutator, and $\Gamma$ is the absorption rate of the sink which is set to $0.5$ in all of the calculations in this paper. 
The quantities $\sigma^{+}_{\rm{f}}$ (${\sigma}^{-}_{\rm{f}}$) are the creation (annihilation) operators at the site connecting site to the sink, and 
 ${\sigma}^{+}_{\rm{sink}}$ (${\sigma}^{-}_{\rm{sink}}$) are the creation (annihilation) operators at the sink.

Once an excitation is initially injected to the initial $i$-th site, we have $\rho(0)=|i\rangle \langle i|$. In order to measure EET from the initial site to the sink, we 
integrate the master equation (\ref{master}) and calculate the population of the sink $\langle {\rm sink}|\rho(t)|{\rm sink}\rangle$ at time $t$. 
The population of each node is $\langle n|\rho(t) |n\rangle$.  
Another quantity of interest is  the \emph{system efficiency} which is defined as the long term sink population:
\begin{equation}
\eta_{\infty}={\rm lim}_{t\rightarrow \infty} \langle {\rm sink}|\rho(t)|{\rm sink}\rangle,
\end{equation}
and determines the fraction of excitation energy transferred into the sink in the long time limit. 

Our calculations are done using the python package  QUTIP ~\cite{qutip} for numerical integration of the Lindblad master equation (\ref{master}), and all energies, time scales, 
and rates are expressed in the units of on-site exception energies $\omega_n$, and since we assume that the network iconsists of the identical units, 
we set $\omega_n = 1$ for all $n$.

\section{Fully Connected Network}
\label{fcn}

\begin{figure*}[t]
\includegraphics[width=0.50\columnwidth]{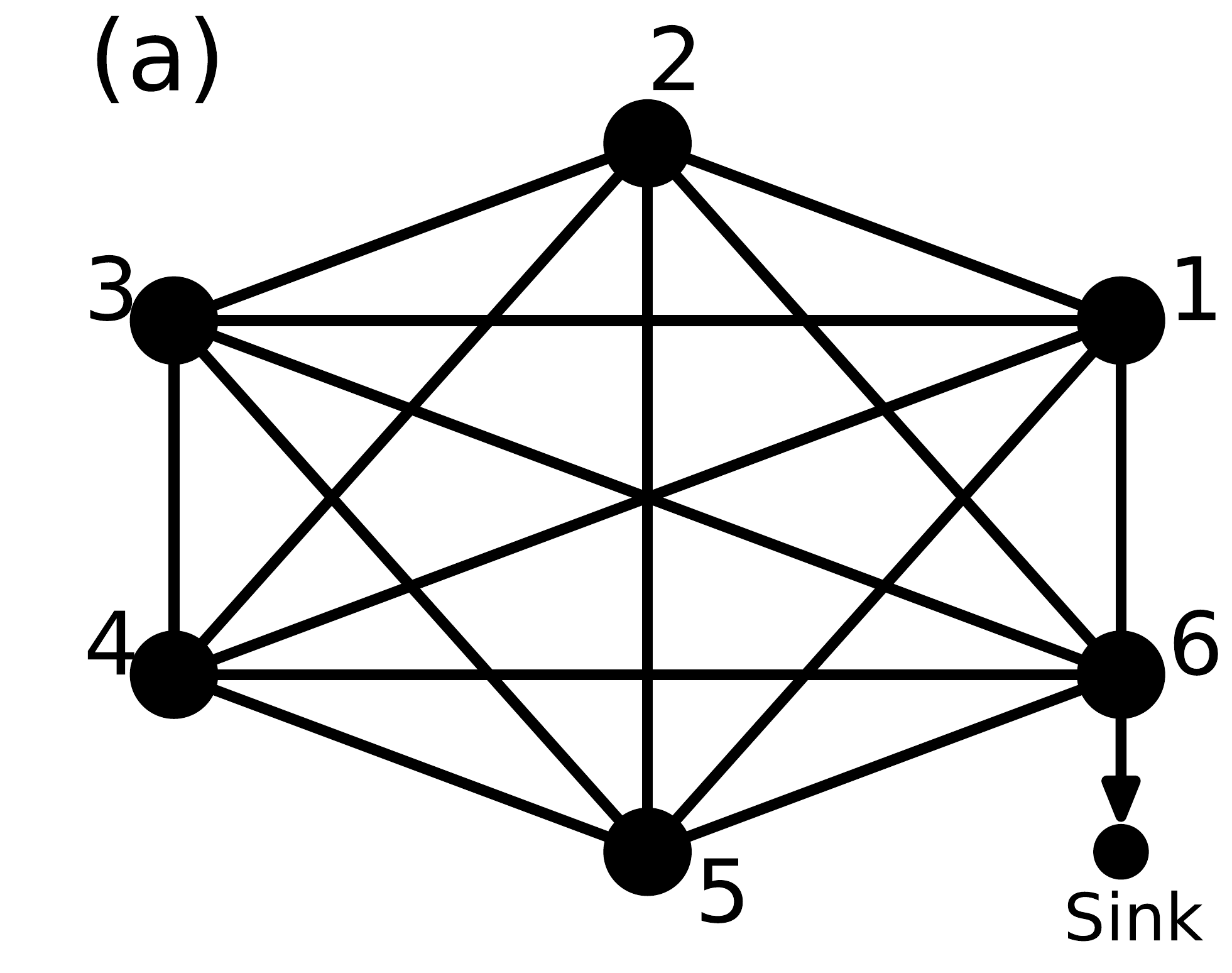}
\includegraphics[width=0.60\columnwidth]{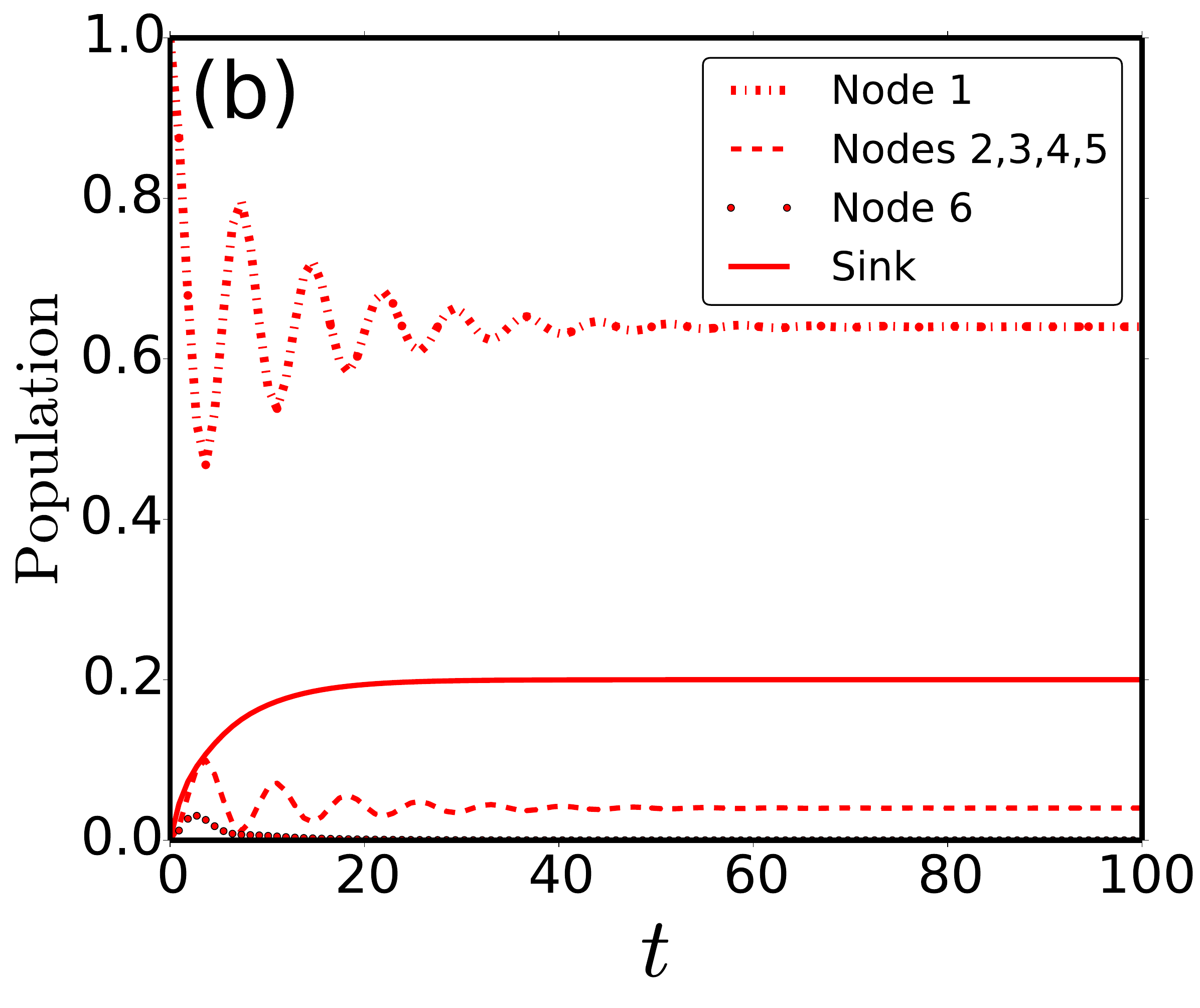}
\includegraphics[width=0.70\columnwidth]{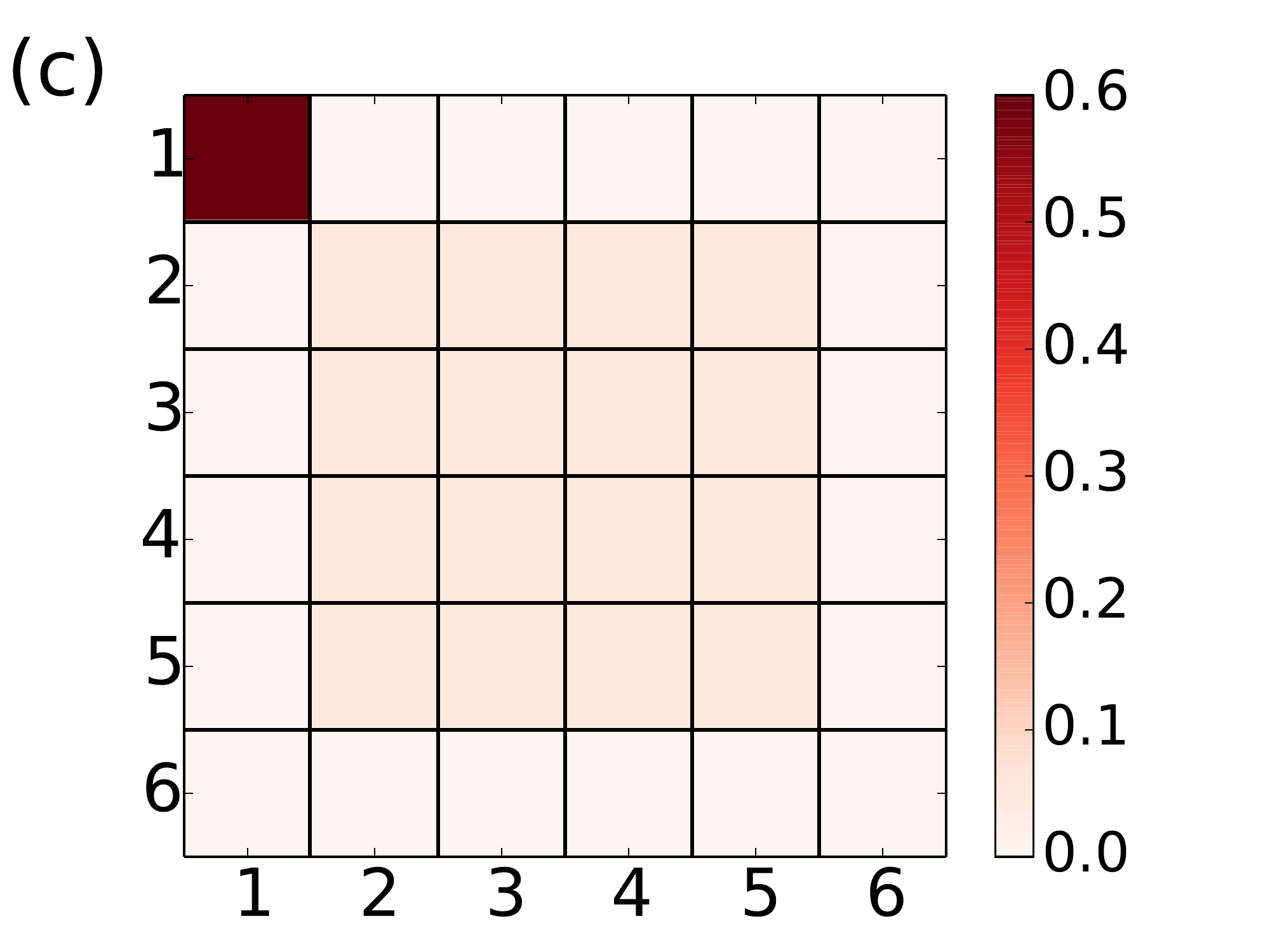}
\caption{ (a) Schematic model of a fully connected six-node network (FCN), where site $6$ is connected irreversibly to the sink. (b) Population at nodes $1, 2,3,4,5$ and $6$ in the network for 
$\Gamma =0.5$. Note that the variation of the populations for the nodes $2,3,4,5$ is exactly the same, as expected from symmetry. 
(c) Density plot of the stationary density matrix $\rho(\infty)$ of the network.}
\label{fcn}
\end{figure*}


We first consider EET in a fully connected network. In Fig.  Fig.~\ref{fcn}(a) we show a fully connected six-site network (FCN). The sink is connected to the node $6$ and the dissipationless 
excitation is injected to 
the node $1$. Figure \ref{fcn}(b) shows the time dependence of the population of the nodes and the sink, calculated using numerical integration of Eq. (\ref{master}).
It can be seen that the system efficiency (sink population) tends to $\approx 0.2$, while most of injected energy ($\approx 80\%$) remains inside the network, mostly on the first node ($\approx 64\%$ of energy remains in node $1$ and $\approx 16\%$ is shared equally among the nodes $2,3,4$ and $5$). This result is a manifestation of the localization of single particle 
states within a fully connected network as already pointed out in Ref. \onlinecite{plenio.The Journal of Chemical Physics2009}.
The reason for localization is the existence of destructive interference of transition amplitudes inside the network. 
To explicitly show this, we expand the initial state  $|1\rangle$ in terms of the orthonormal eigenstates of the tight-binding Hamiltonian (\ref{TBH}), resulting in:
%
\begin{eqnarray}
|1\rangle =&&-0.71 \left( -0.71, 0.71, 0, 0, 0, 0\right)\nonumber  \\
                 && -0.41 \left(  -0.41, -0.41,  0.82,  0, 0, 0 \right)\nonumber\\
                 &&-0.29 \left( -0.29, -0.29,  -0.29,  0.87, 0, 0 \right)\nonumber\\
                 &&-0.22 \left( -0.22, -0.22,  -0.22,  -0.22, 0.89, 0 \right)\nonumber\\
                 &&-0.18 \left(-0.18, -0.18,  -0.18,  -0.18, -0.18, 0.91 \right)\nonumber\\
                 &&+0.41 \left(0.41, 0.41,  0.41,  0.41, 0.41, 0.41 \right).
\label{fcn-exp}
\end{eqnarray}
%
Eq. (\ref{fcn-exp}) clearly shows that in the four out of six terms in the expansion of the initial state, the sixth node (which connects the network to the sink) has no contribution, 
and hence most of the injected energy cannot reach this site and be transferred to the the sink. For visualization of the energy localization, 
the long time limit of the density matrix of the network is illustrated in Fig.~\ref{fcn}(c), where the block diagonalization of the stationary density matrix  is an indication of energy localization in the network.  

\begin{figure}[t]
\includegraphics[width=0.8\columnwidth, height=0.7\columnwidth]{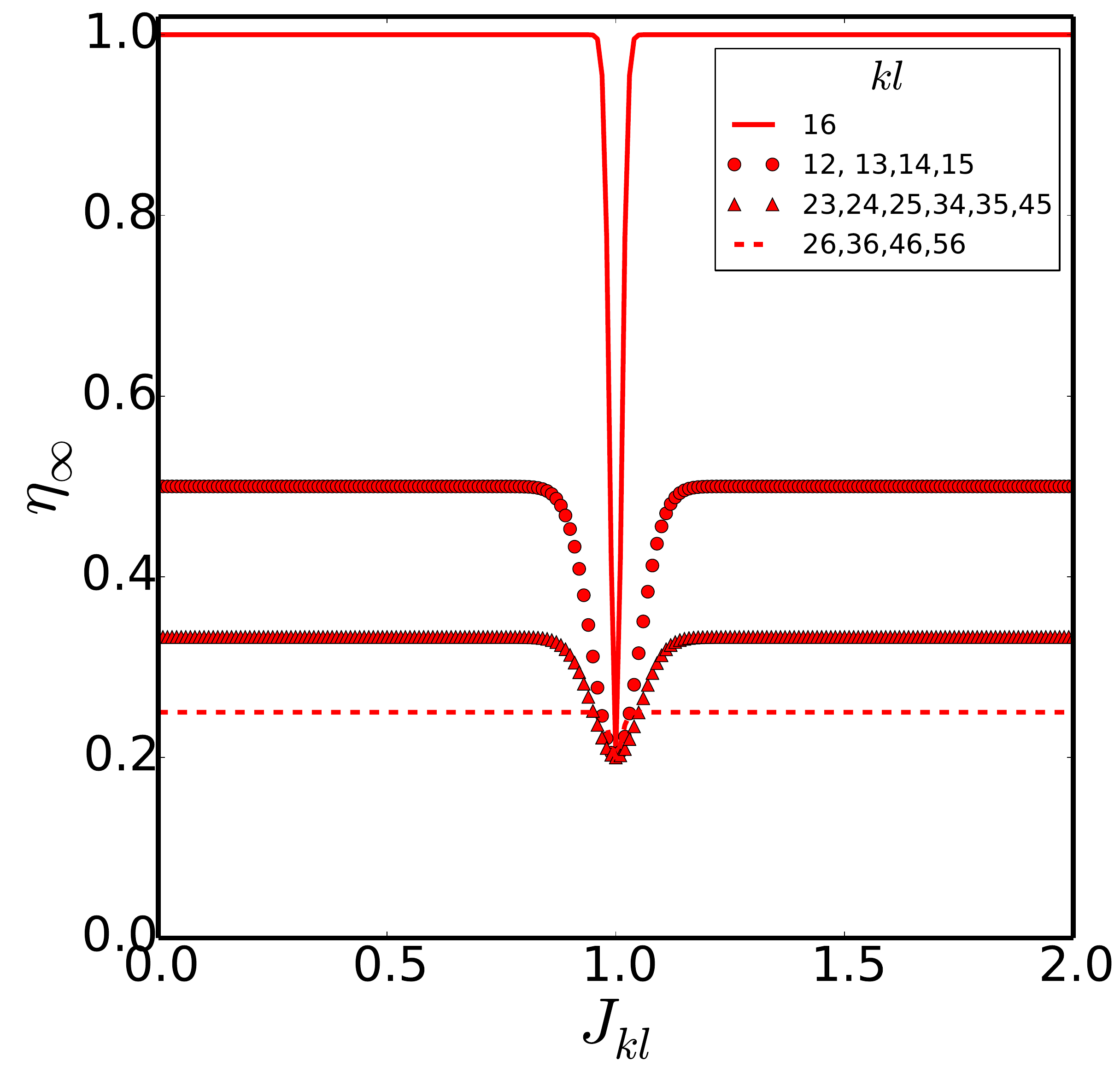}
    \caption{System efficiency as a function of the value of the the hopping integral between the different nodes $l$ and $k$ ($J_{lk}$). It can be seen that the hopping rate $J_{16}$ connecting
the insertion point and the sink has the strongest influence on the efficiency. }
    \label{pop-J16}
\end{figure}

To investigate the effect of symmetry on the energy  localization, we calculate the system efficiency when the hopping integral between the nodes $1$ and $6$ varies, 
while the rest of hopping integrals $J_{nm}$ ($m,n \neq 1, 6$) remain equal to unity.  The result is displayed in Fig.~\ref{pop-J16} and it shows that the system efficiency 
($\eta_{\infty}$) is highly sensitive to the value of $J_{16}$. For $J_{16}=1$, $\eta_{\infty}$ is minimized, but when the hopping integral between 
the two nodes $1$ and $6$ slightly deviates from the other ones, the system efficiency rapidly rises to unity. 
This result shows that the state localization inside a fully connected network is highly sensitive to the symmetry of the Hamiltonian and any asymmetry 
due to $J_{16}$ destroys the destructive interference loops and hence localization in the network. 
The important conclusion here is that such a symmerry breaking dramatically increases the efficiency of EET.  Figure \ref{pop-J16} also shows that introducing asymmetry into
the other links slightly increases the system efficiency, but does not promote the system to become a perfect transmitter.


\section{Influence of edge deletion on system efficiency}
\label{edge-deletion}

\begin{figure}[t]
\includegraphics[width=0.35\columnwidth]{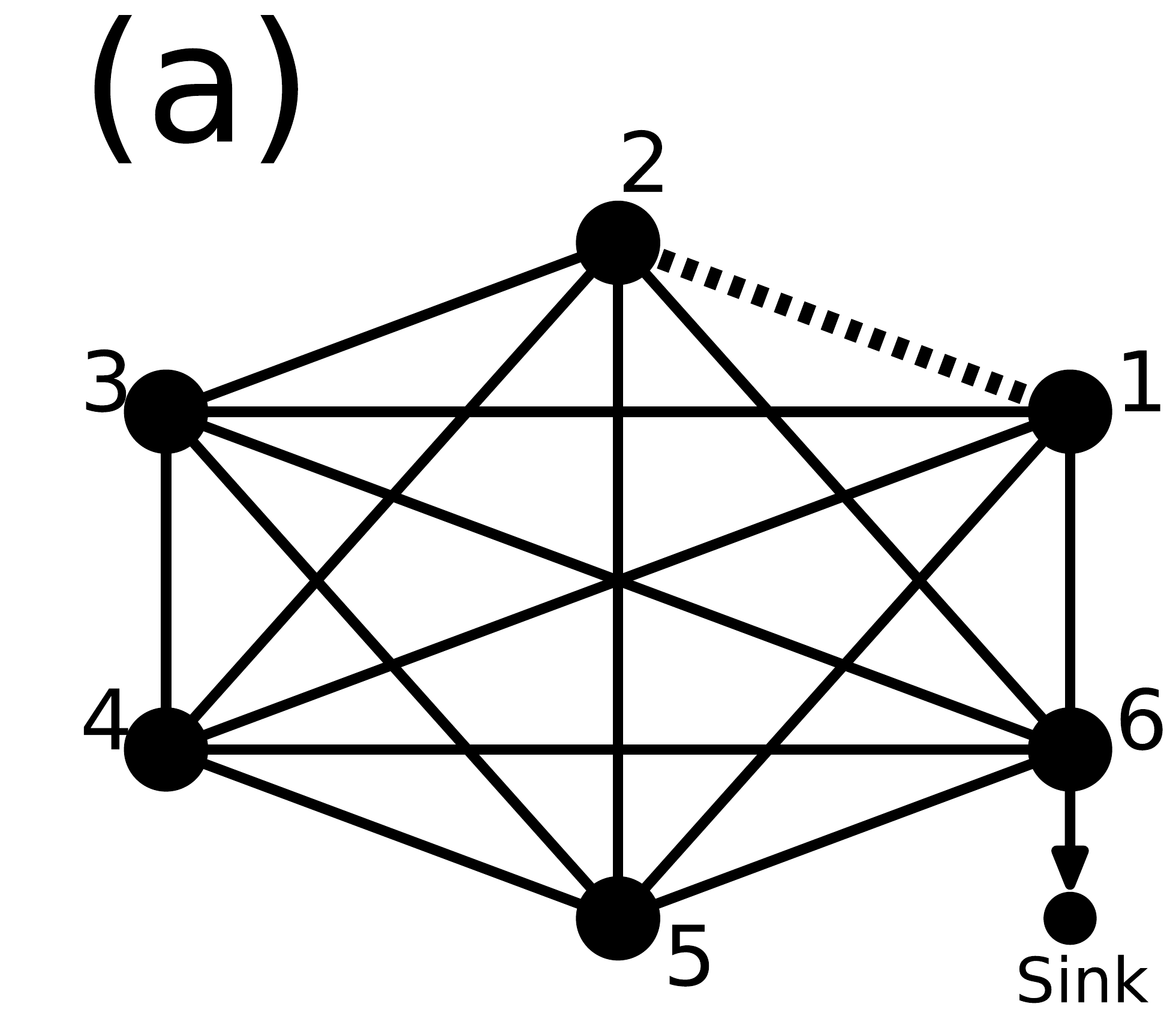}\includegraphics[width=0.4\columnwidth]{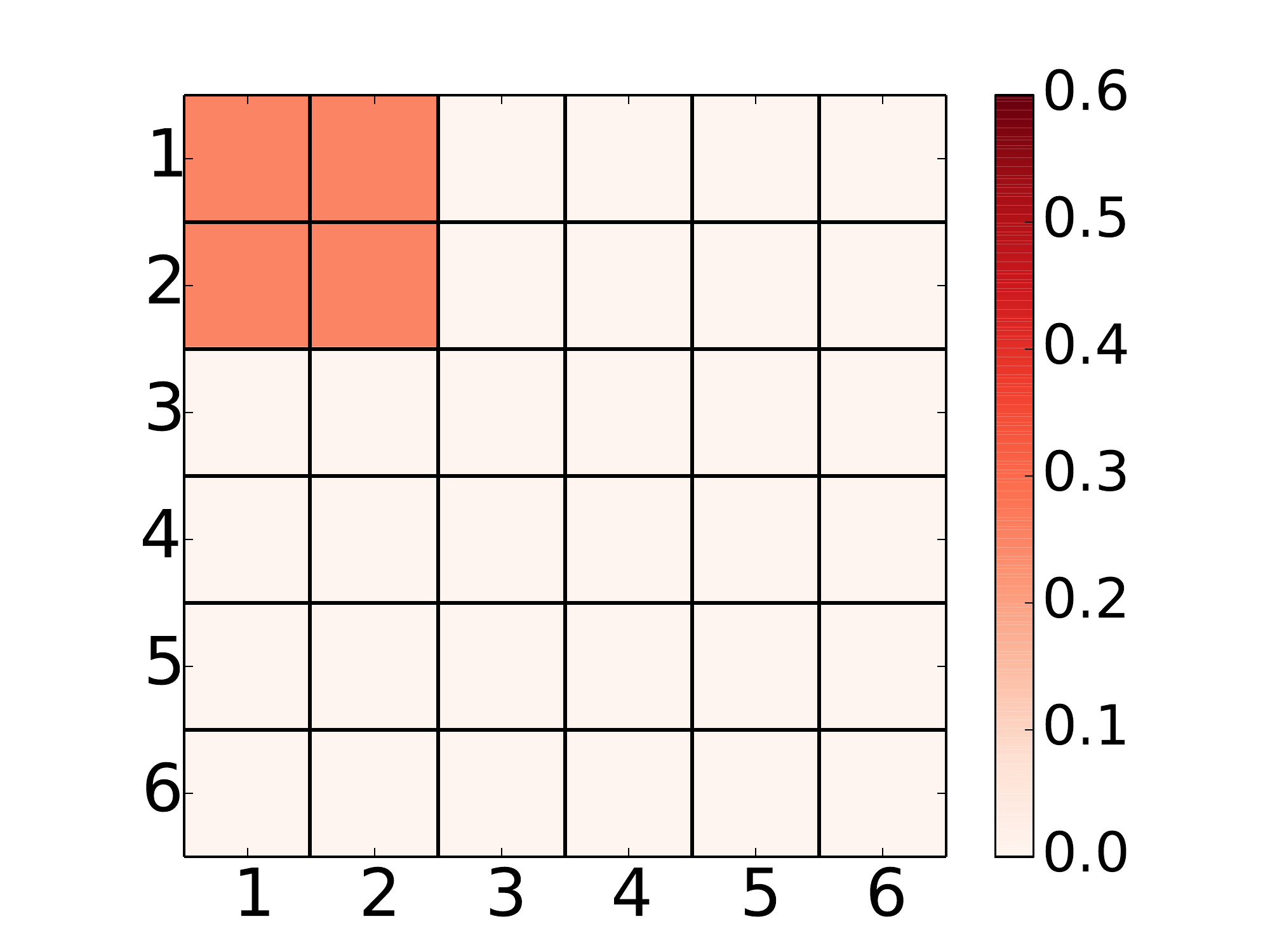}
\includegraphics[width=0.35\columnwidth]{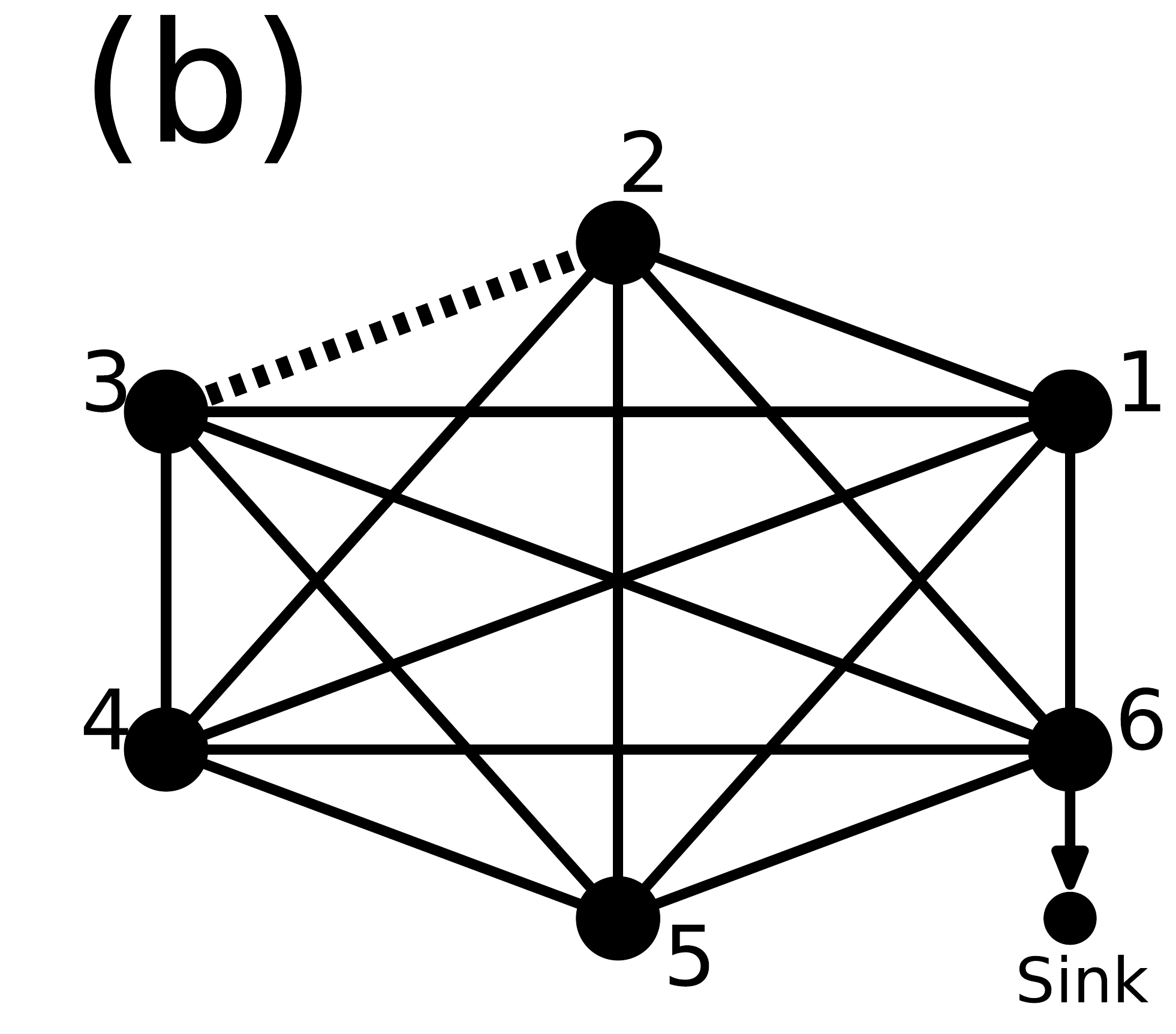}\includegraphics[width=0.4\columnwidth]{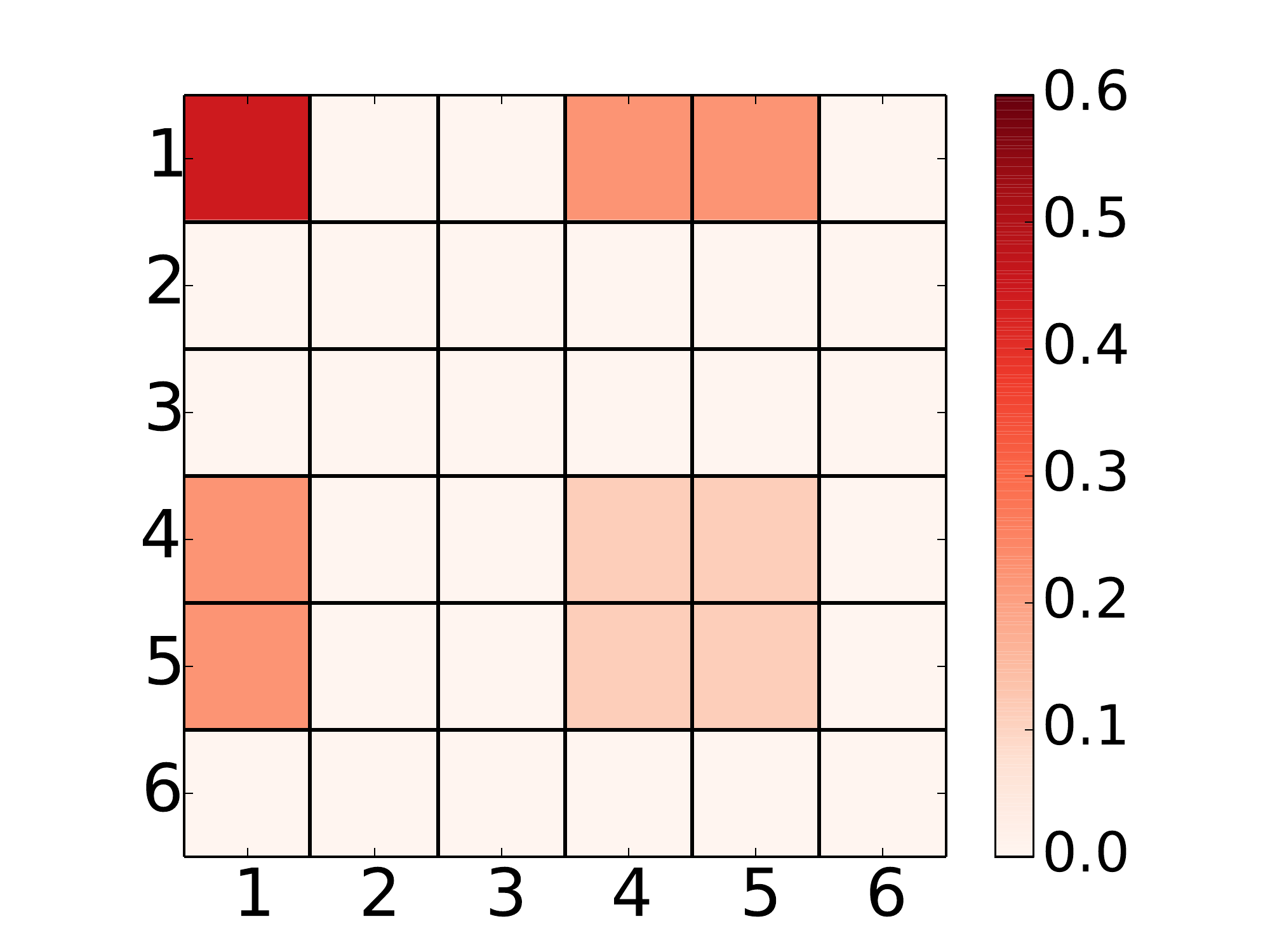}
\includegraphics[width=0.35\columnwidth]{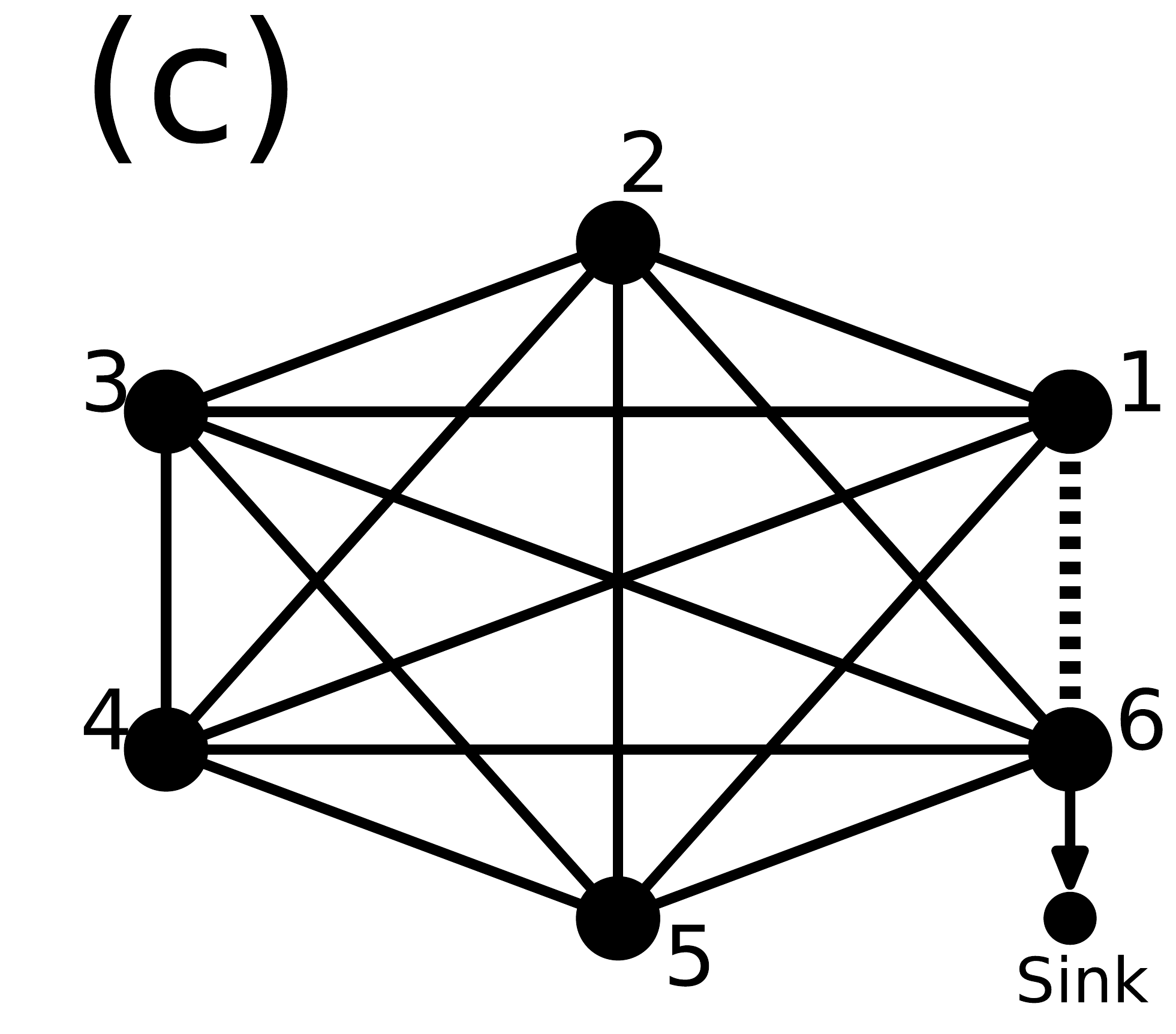}\includegraphics[width=0.4\columnwidth]{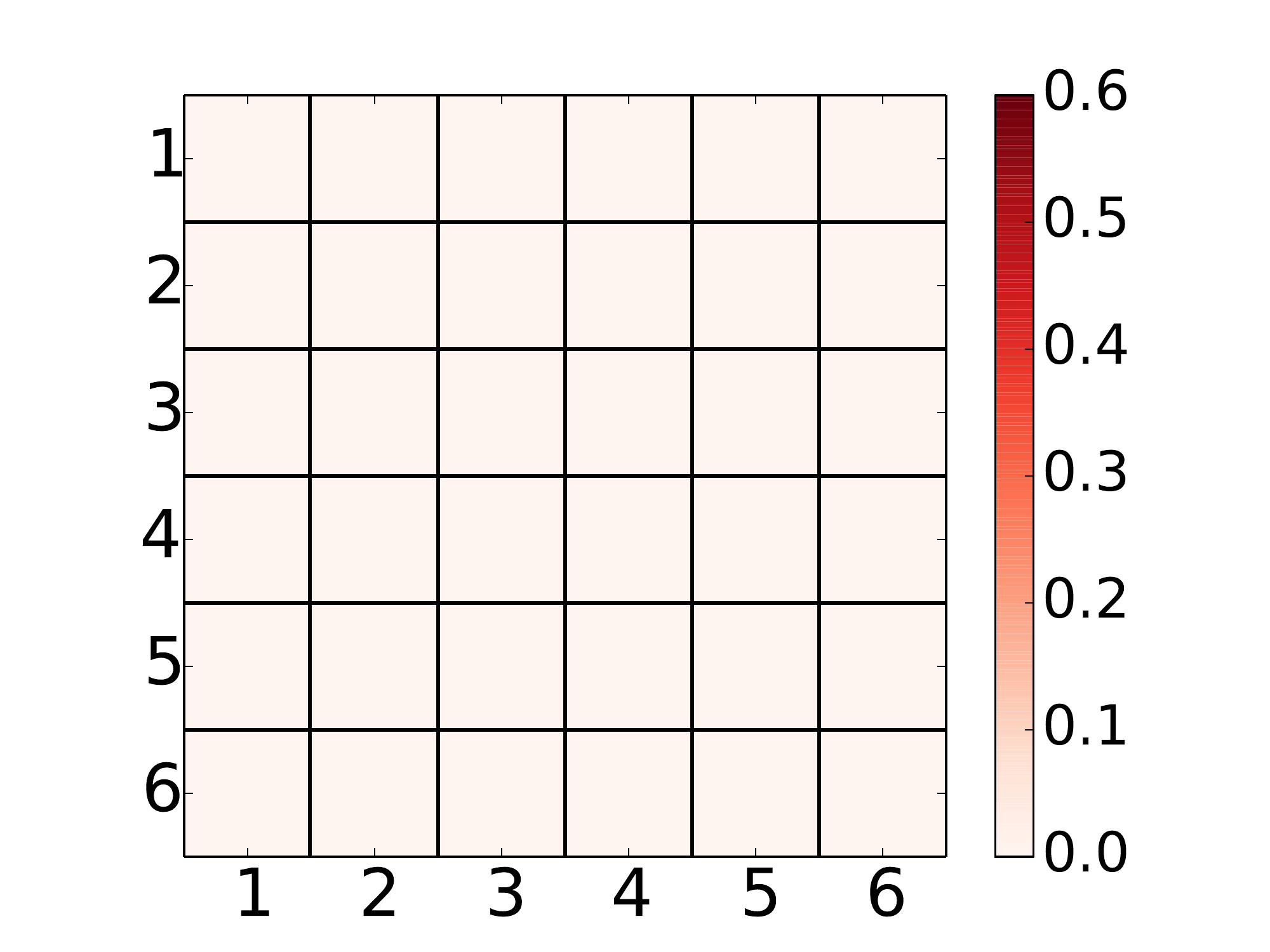}
\caption{(Color online) Density plots showing the behaviour of the elements of the steady state density matrix,  $\rho(\infty)$,  in the case of deletion of edges between the nodes (a) $1, 2$; (b) $2, 3$ and (c) $1, 6$. The dotted lines in the left panels denote the edges deleted. As can be seen in case (c), localization is absent in the network and therefore the EET efficiency to the sink becomes unity.}
\label{edge-delet1}
\end{figure}

\begin{figure*}[t]
\includegraphics[width=0.3\columnwidth]{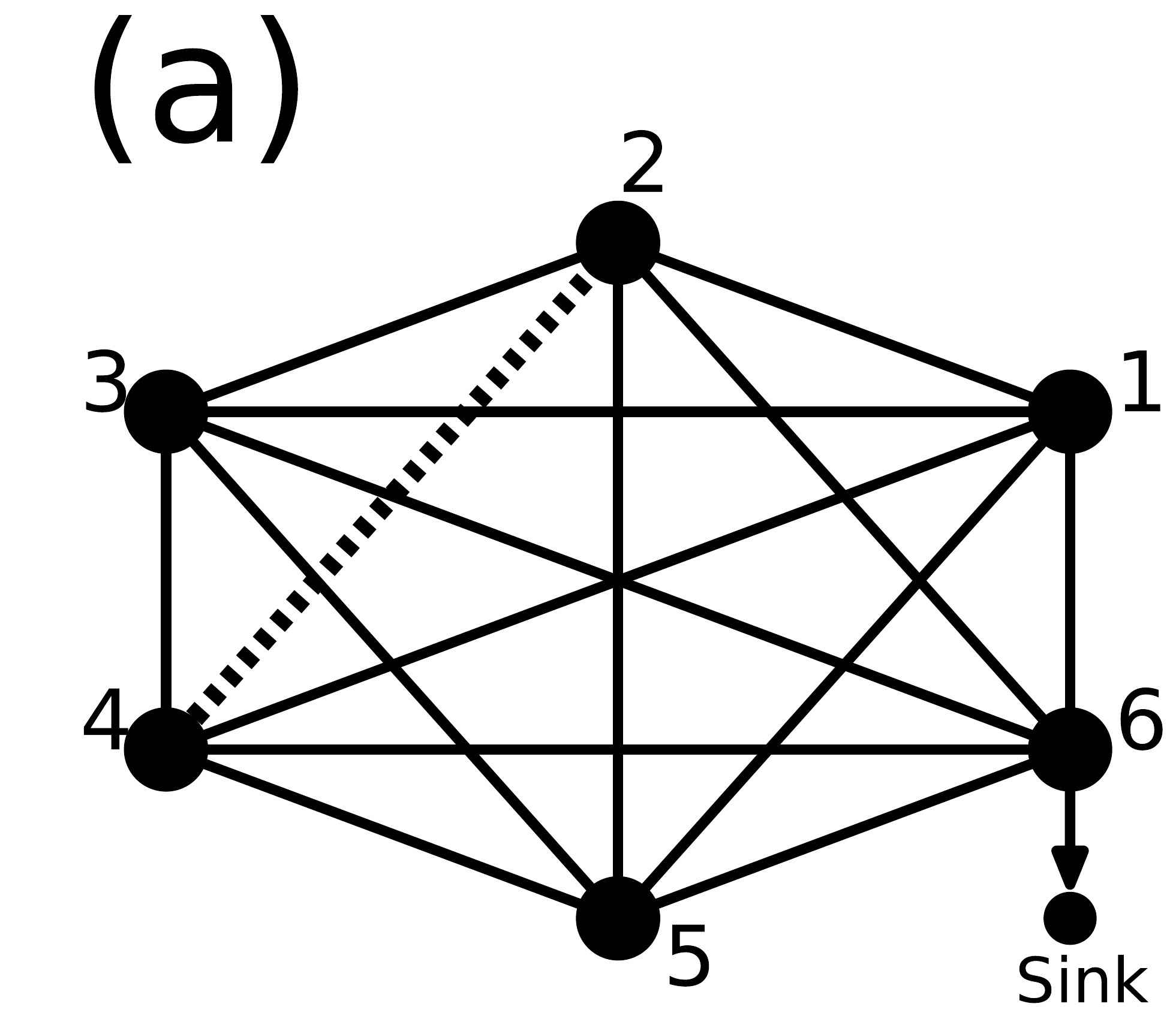}\includegraphics[width=0.35\columnwidth]{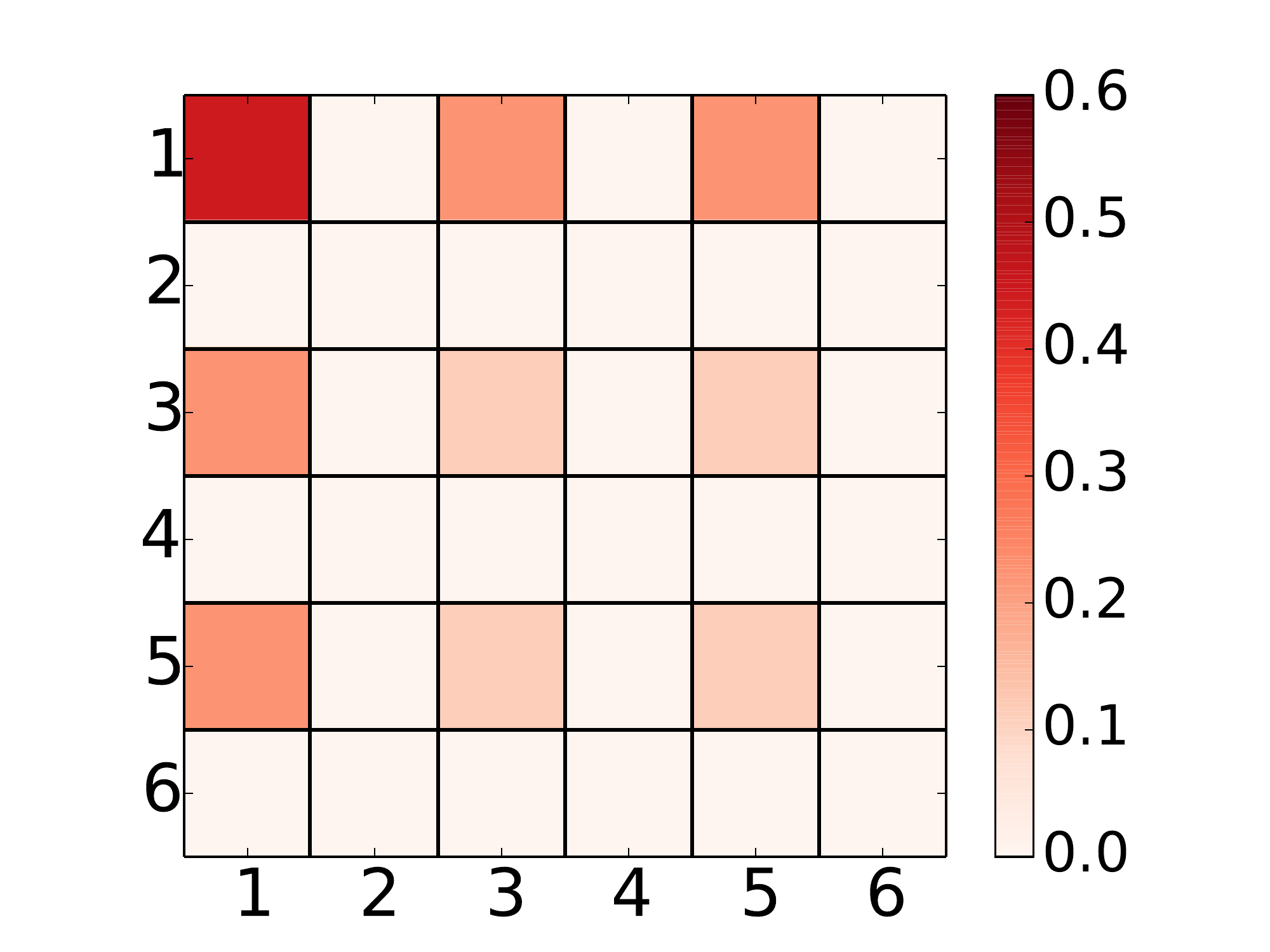}
\includegraphics[width=0.3\columnwidth]{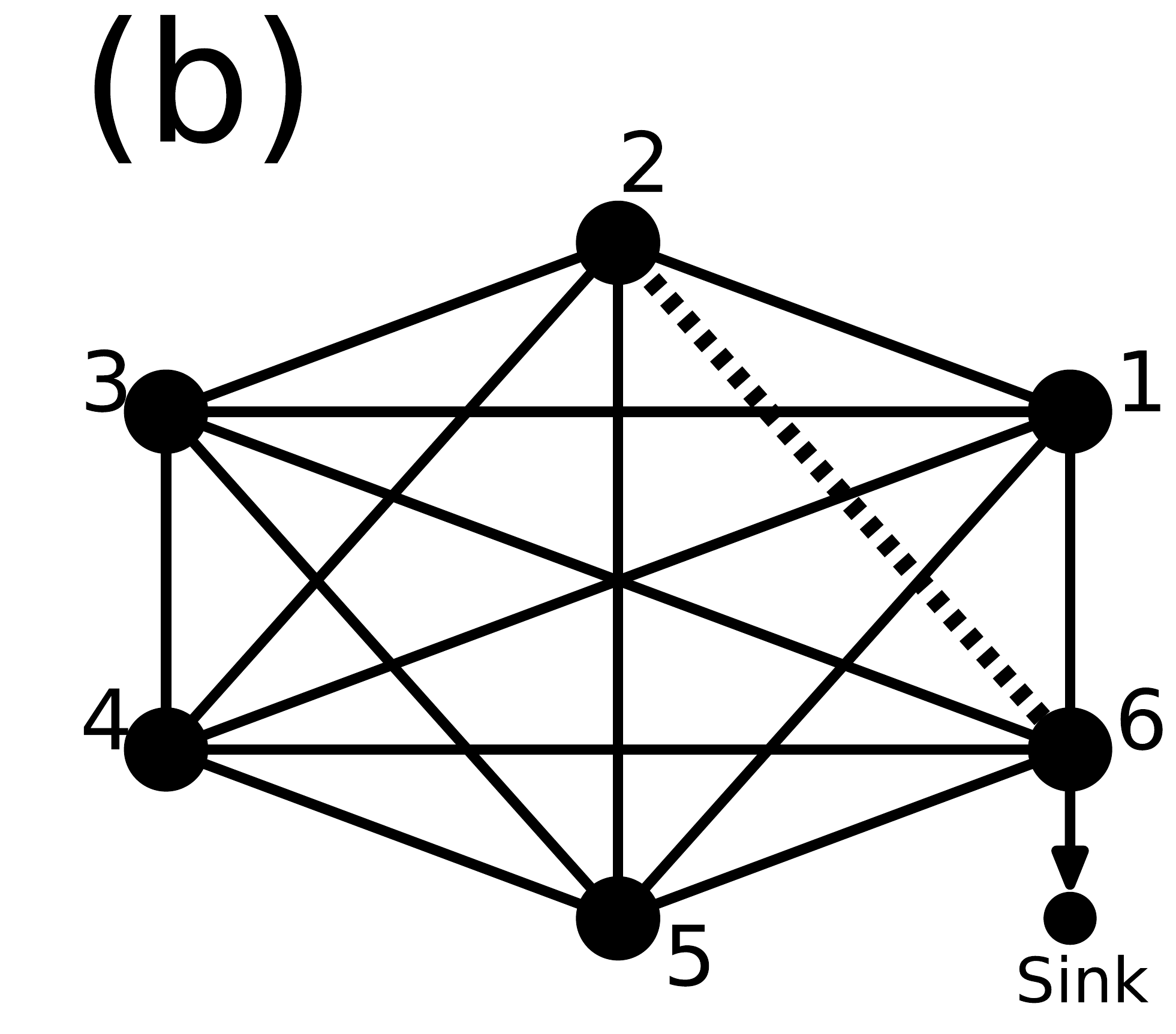}\includegraphics[width=0.35\columnwidth]{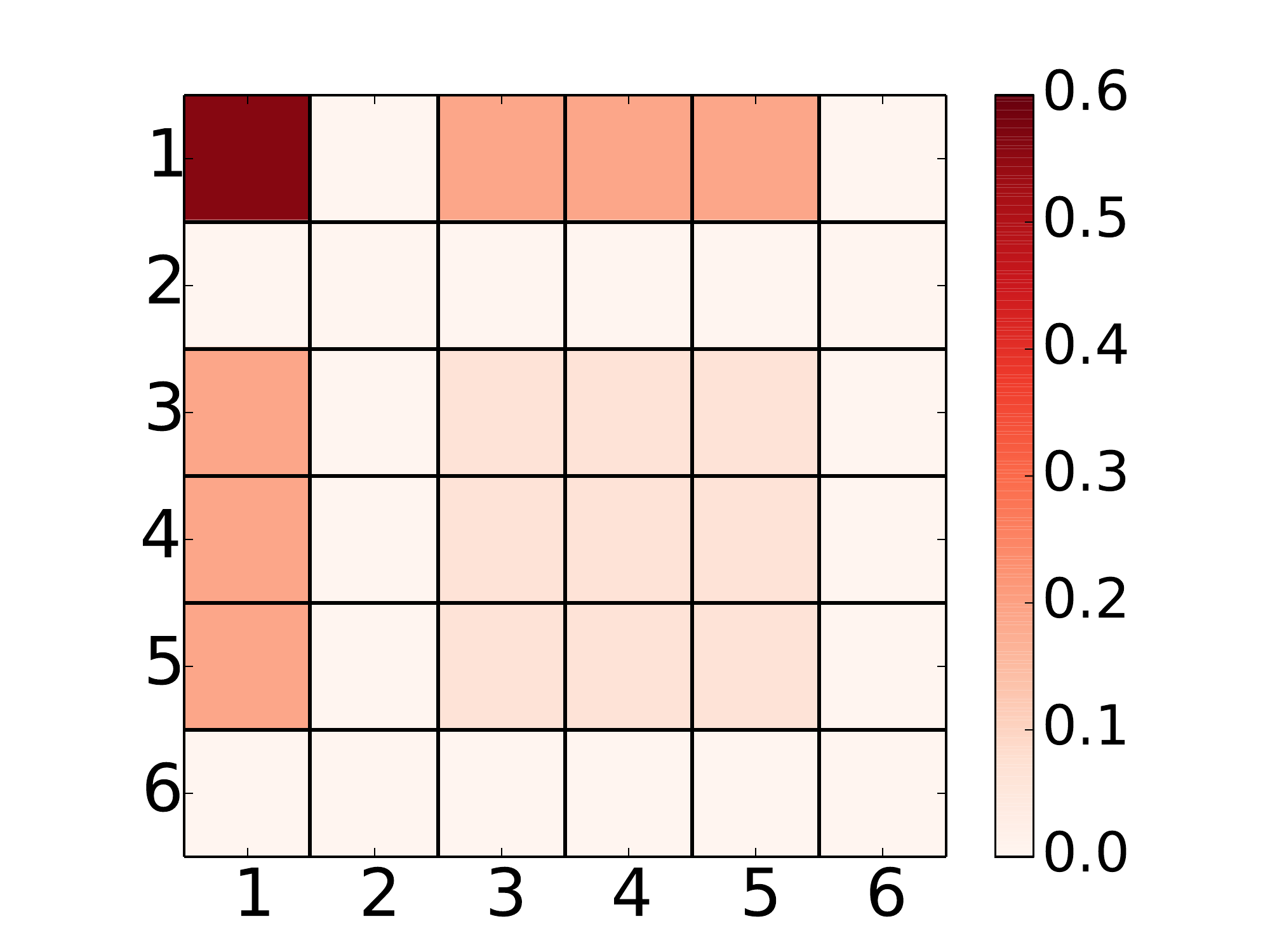}
\includegraphics[width=0.3\columnwidth]{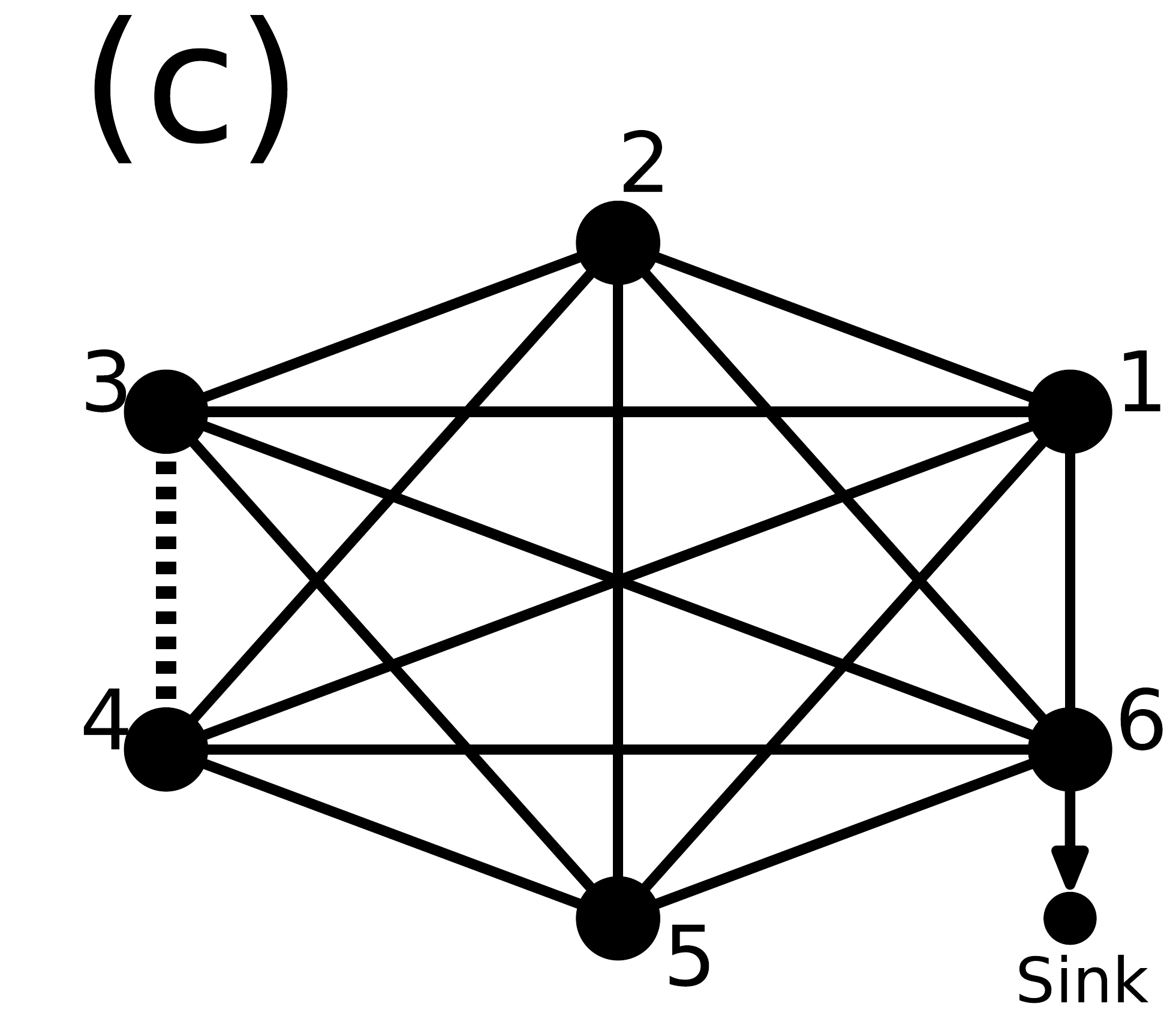}\includegraphics[width=0.35\columnwidth]{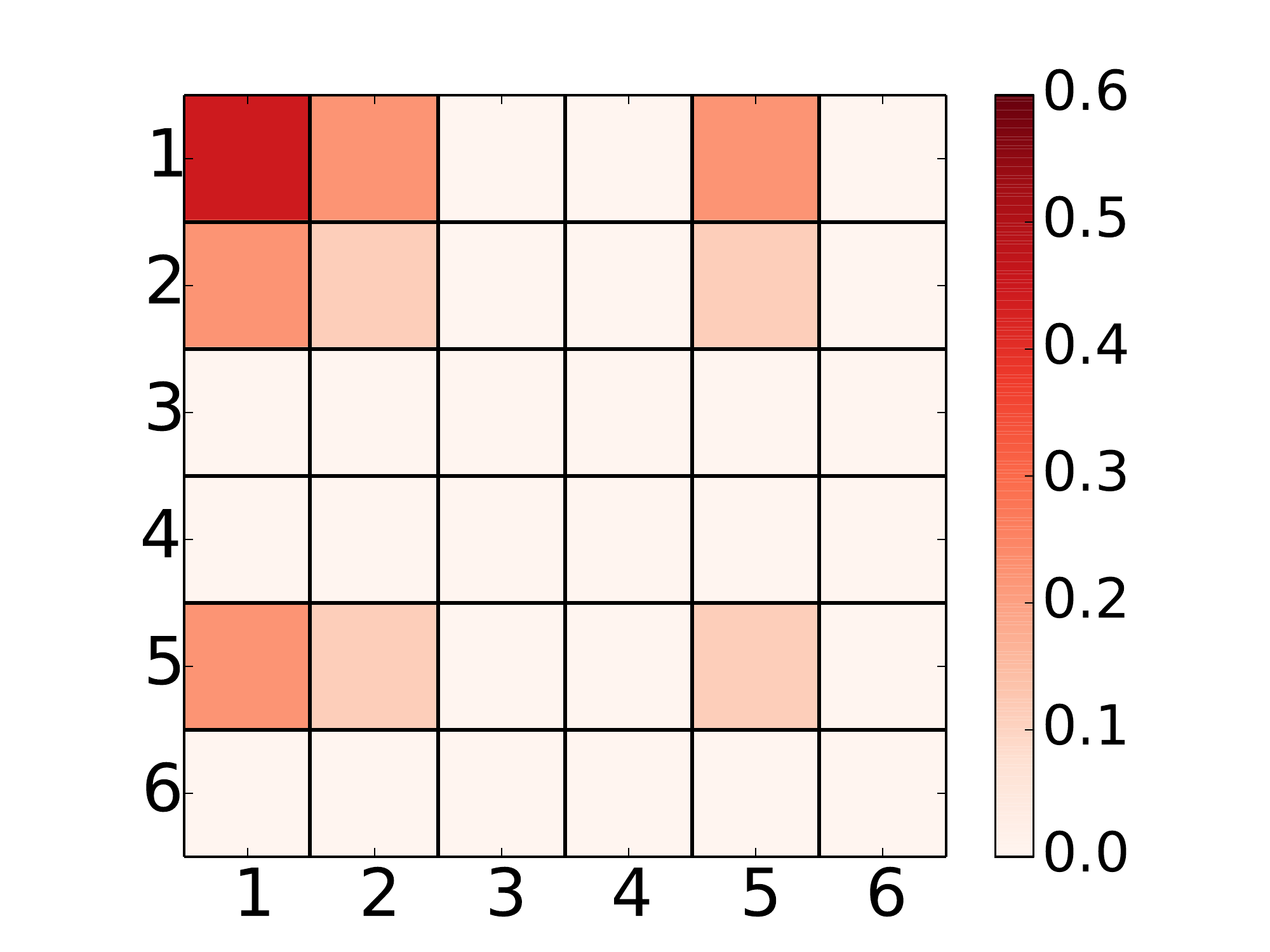}
\includegraphics[width=0.3\columnwidth]{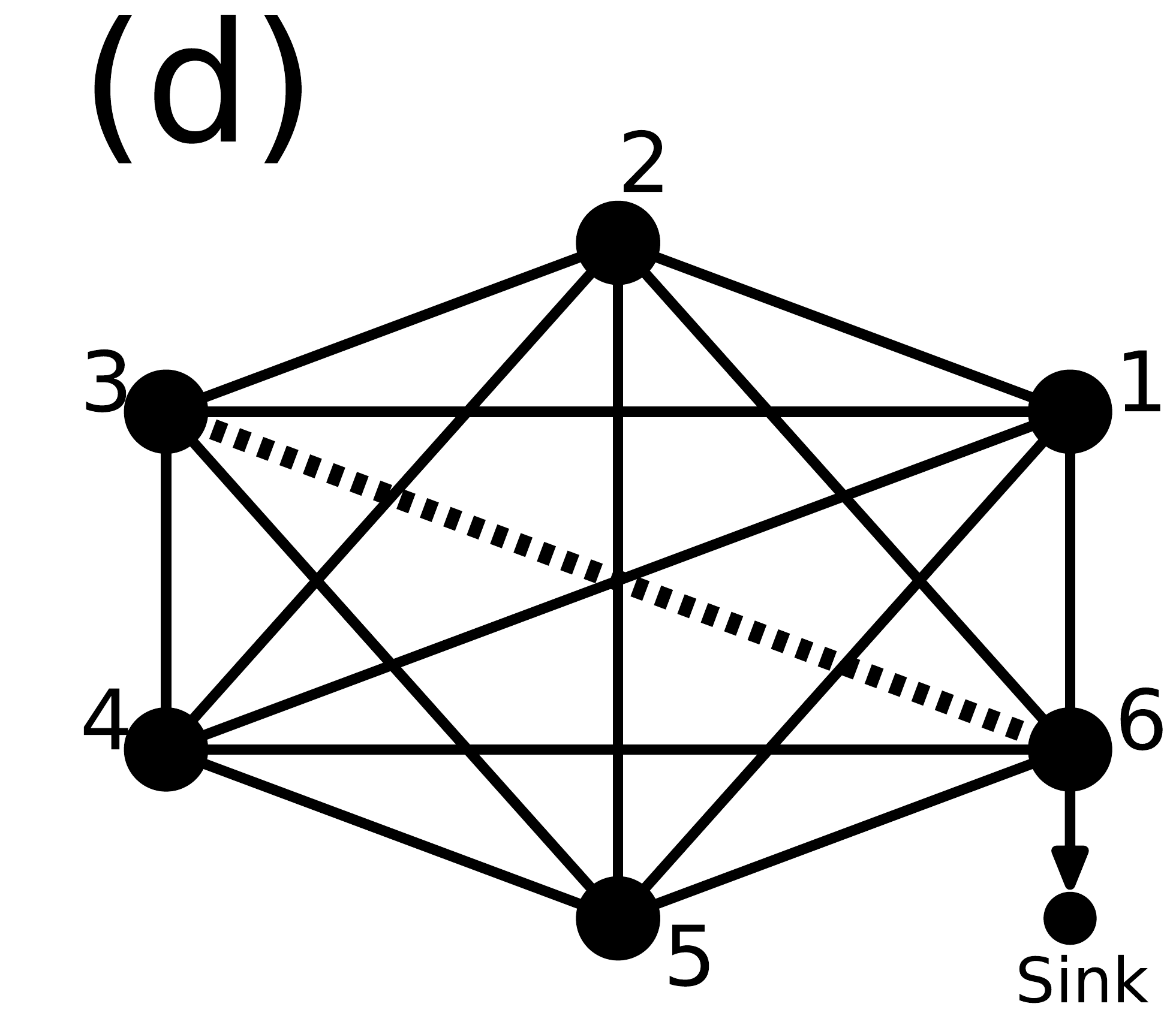}\includegraphics[width=0.35\columnwidth]{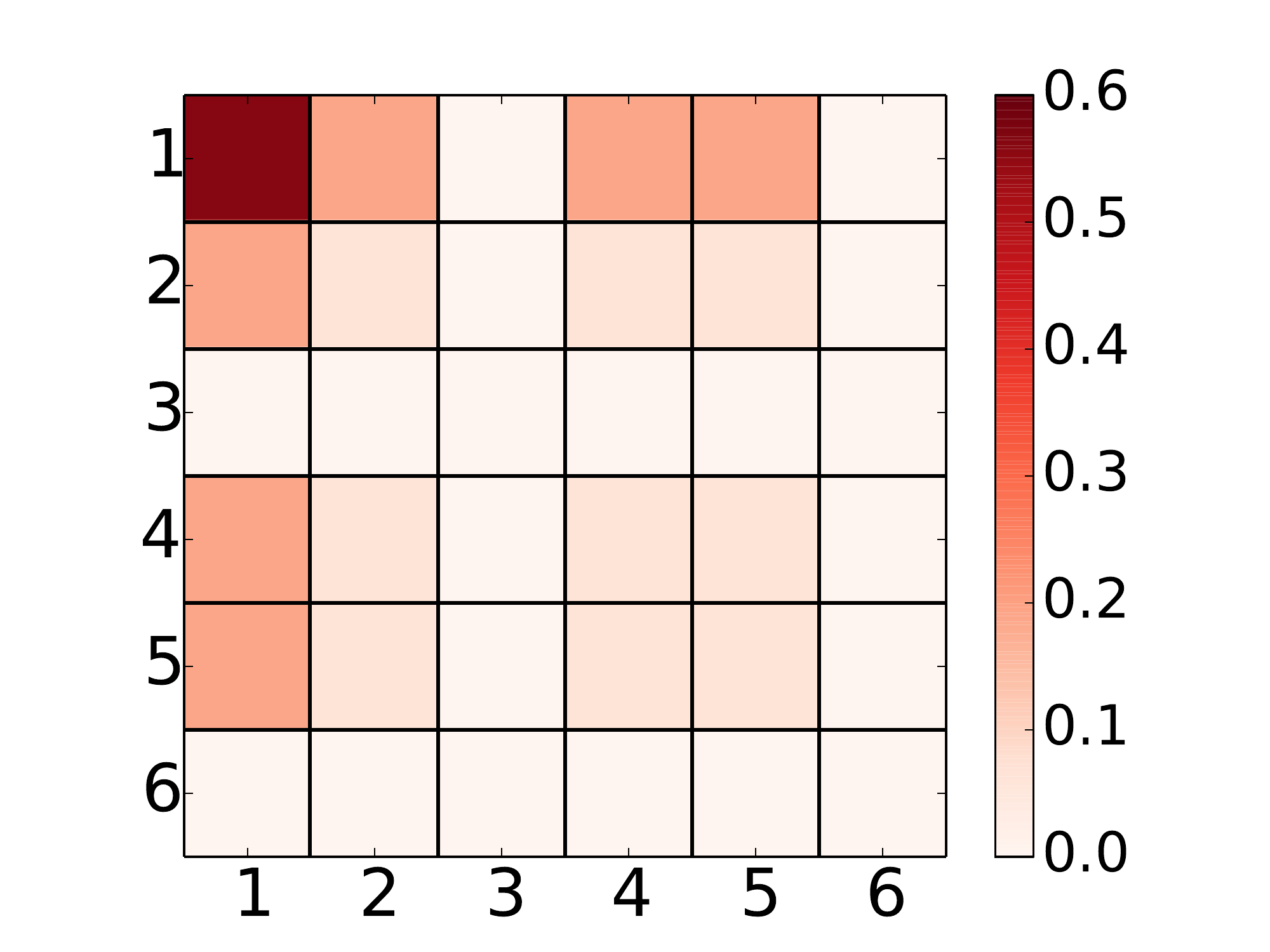}
\includegraphics[width=0.3\columnwidth]{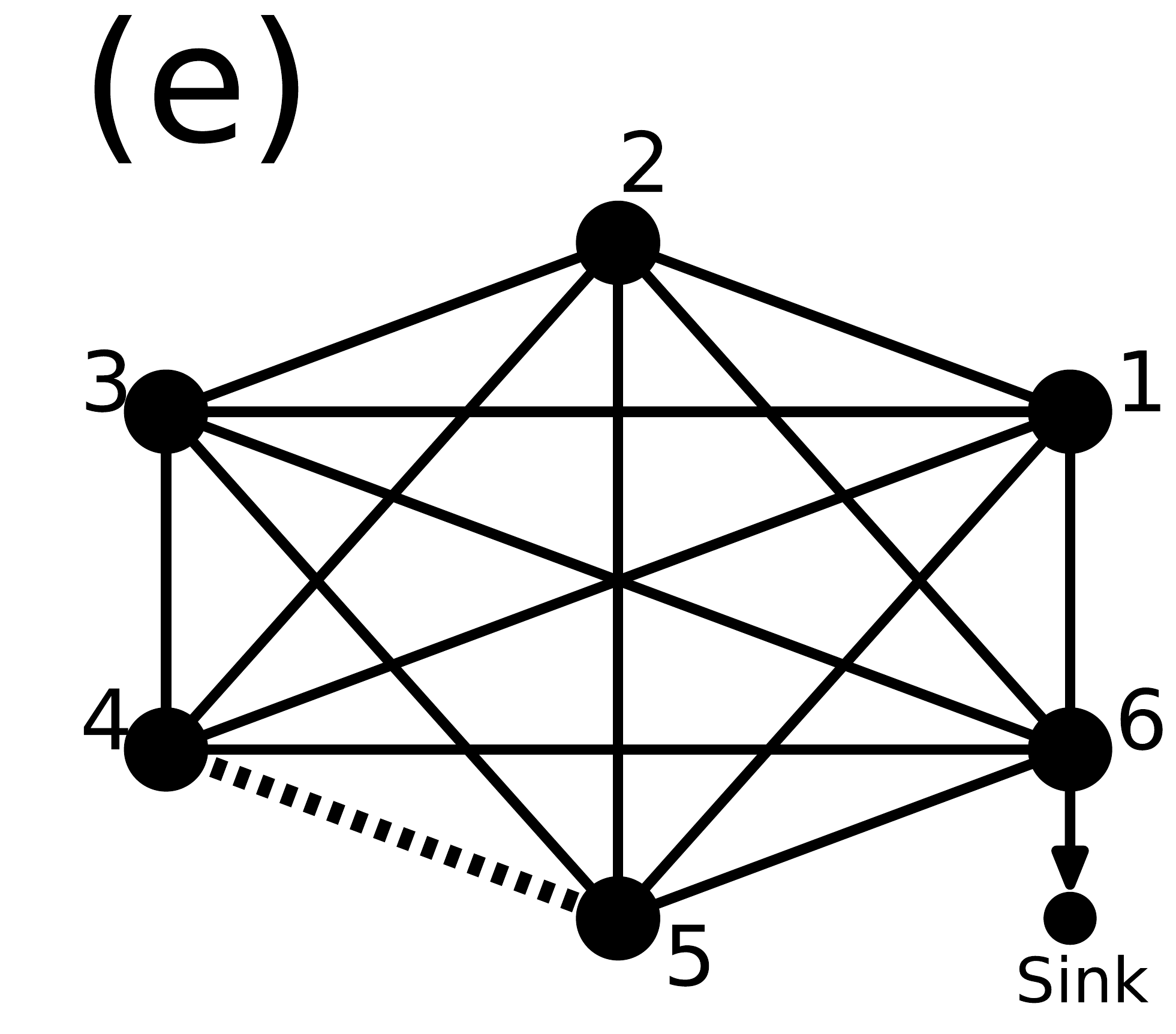}\includegraphics[width=0.35\columnwidth]{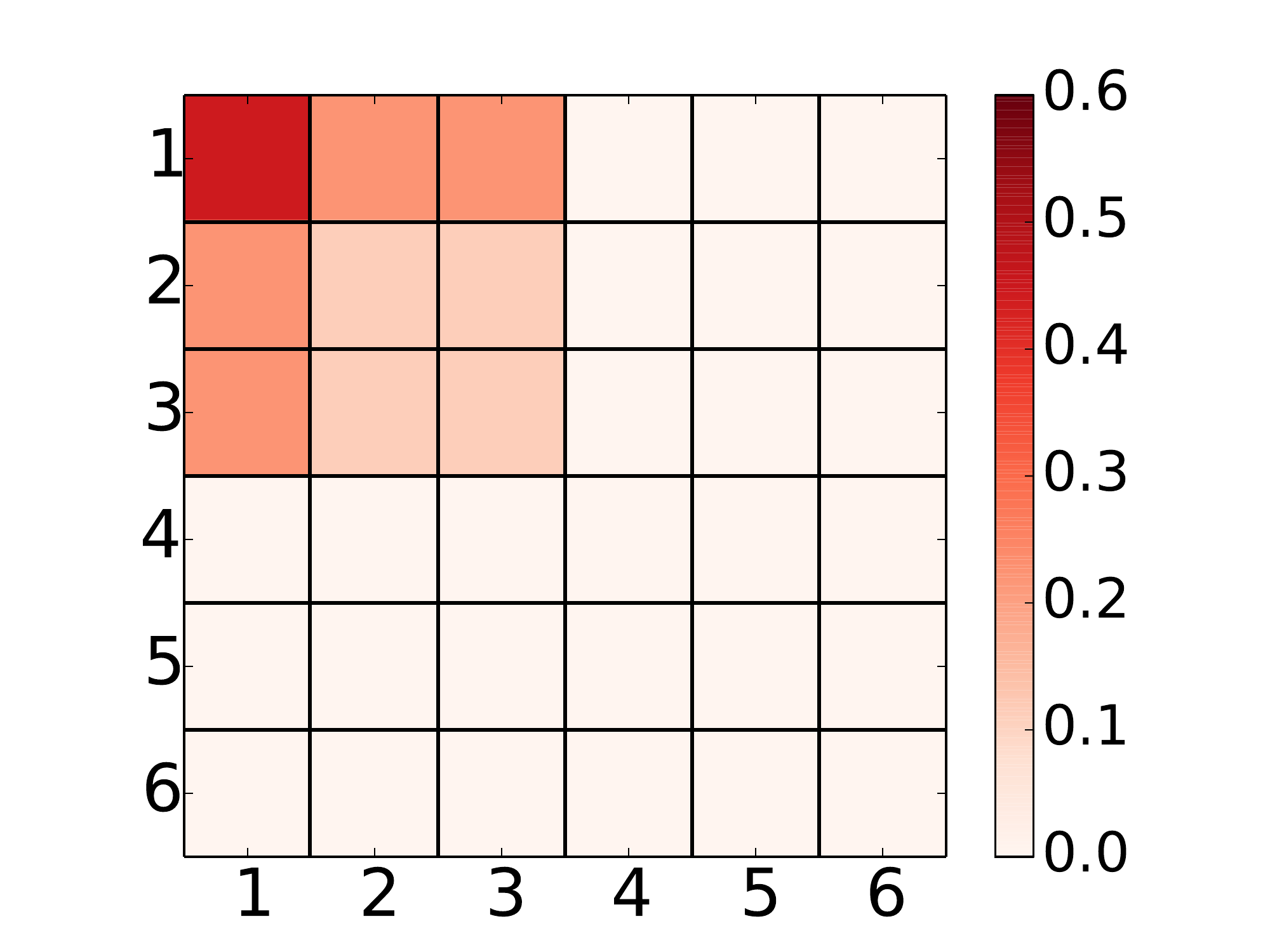}
\includegraphics[width=0.3\columnwidth]{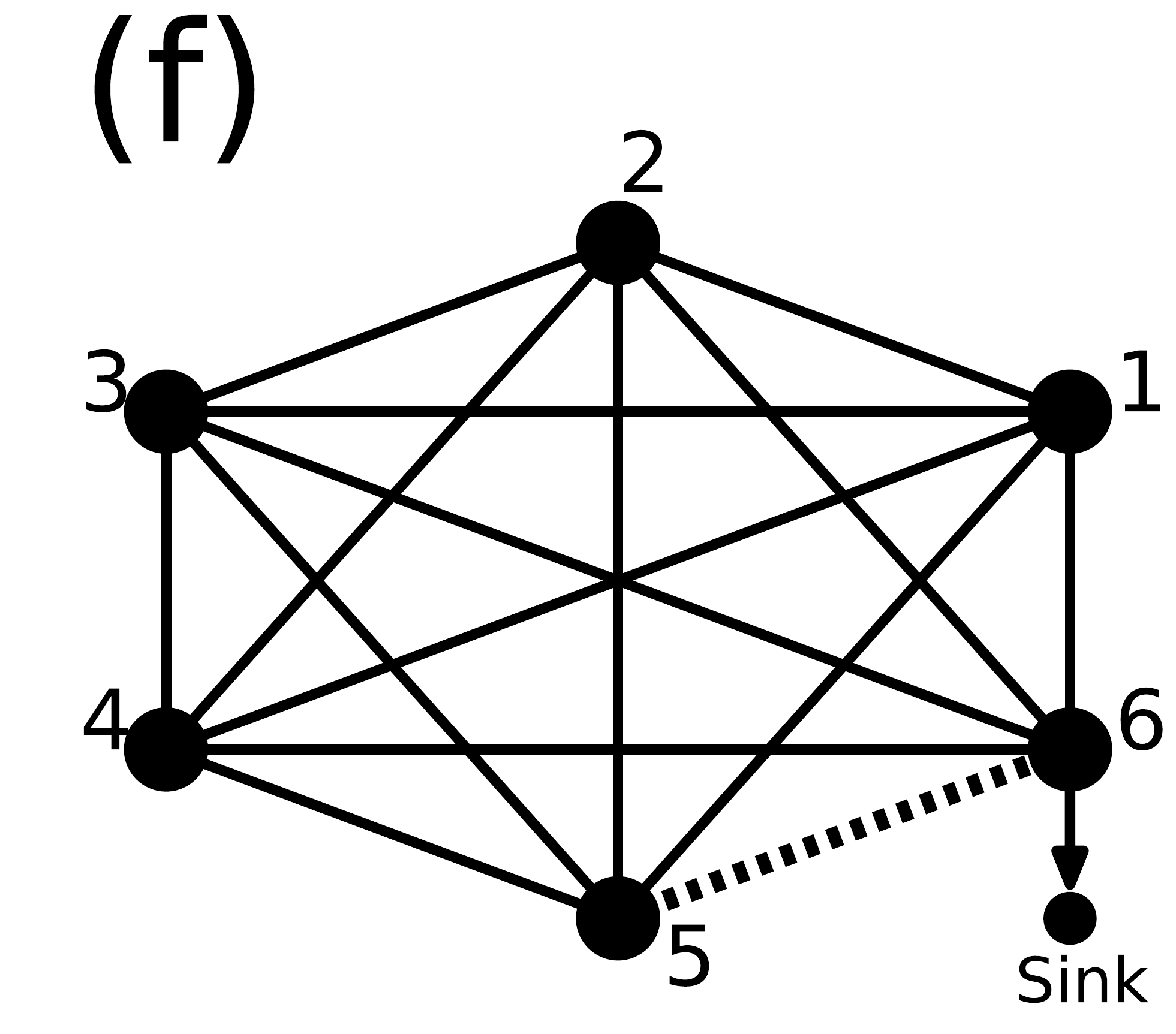}\includegraphics[width=0.35\columnwidth]{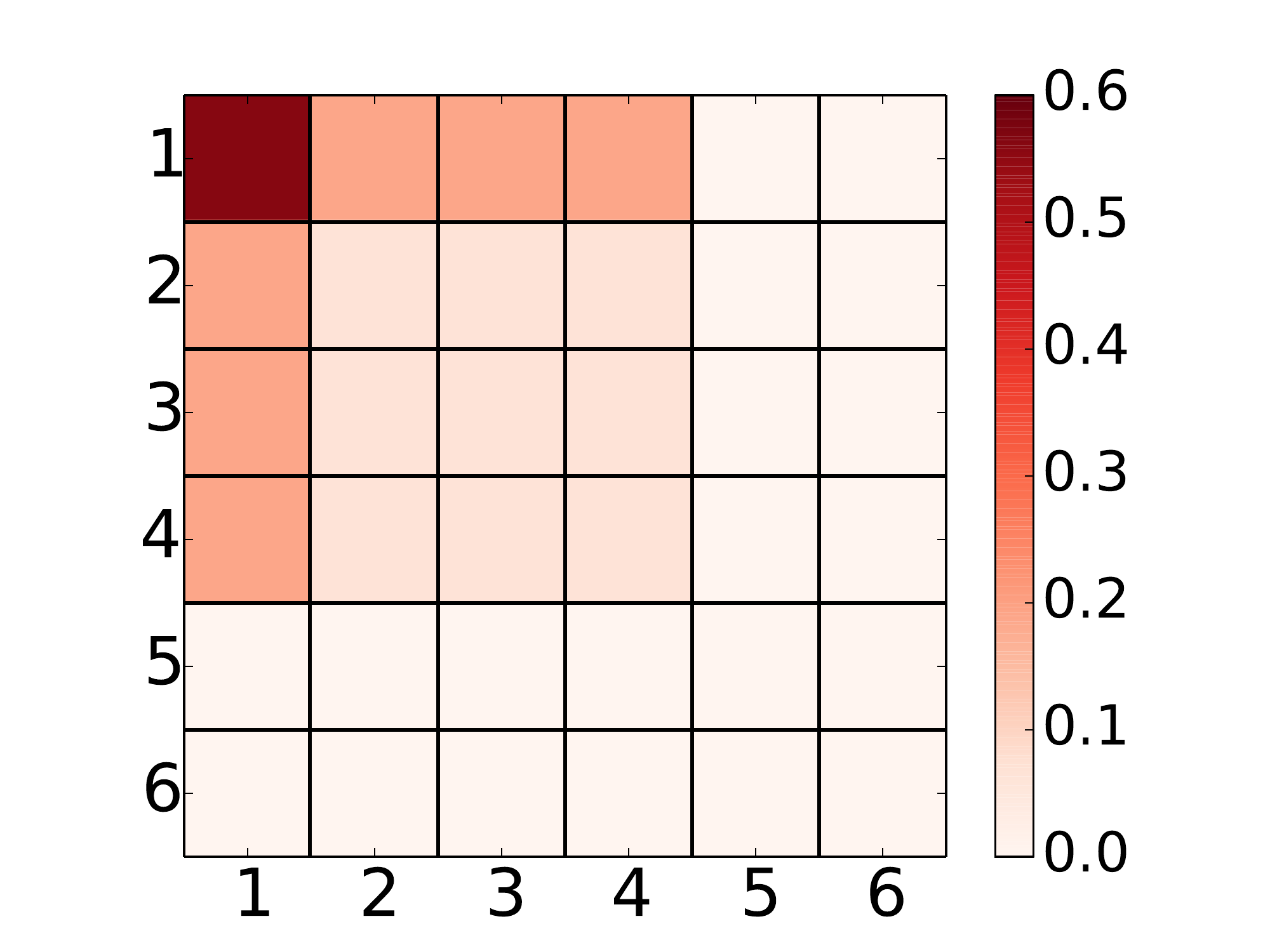}
\caption{(Color online) Density plots showing the behavior of the elements of the steady state density matrix,  $\rho(\infty)$,  
in the case of deletion of edges between nodes (a) $2, 4$, (b) $2, 6$, 
(c) $3, 4$, (d) $3, 6$, (e) $4, 5$, and (f) $5, 6$. The dotted lines in the left panels denote the edges deleted. 
In all the density plots above there is localization and therefore the EET efficiency is low.}
\label{edge-delet2}
\end{figure*}

In the previous section we demonstrated that even a small asymmetry in the hopping rates is able to destroy the state localization in a fully connected network, and hence it increases the energy transfer to the sink. 
Next we proceed to investigate the influence of a defect on energy transport. To this end, we consider cutting one link between any two nodes. Deletion of an edge between the 
nodes means blocking the hopping of the excitation between them. 
Figure \ref{edge-delet1} illustrates the density matrices of the networks with deletion of the links between the nodes $1, 2$ (Fig.~\ref{edge-delet1}(a)), 
$2, 3$ (Fig.~\ref{edge-delet1}(b)) and $1, 6$ (Fig.~\ref{edge-delet1}(c)). It can be seen from this figure that only in the case of deletion of the link between the nodes $1$ and $6$, 
the localization vanishes and the injected energy gets completely transferred into the sink. 
Preventing hopping between the nodes $1, 2$ and $2, 3 $ results in block diagonalization of the steady density matrix and hence localization inside the network is preserved.  
This can be verified by expansion of the initial state in terms of the eigenstates of the Hamiltonian in Eq.(\ref{TBH}), 
which results in the following expression:   
\begin{eqnarray}
|1\rangle =&&-0.707 \left(  -0.71,  0.71,  0,  0, 0, 0 \right)\nonumber\\
                 &&+0.365 \left(  0.36,  0.36,  0.43,  0.43, 0.43, 0.43  \right)\nonumber\\
                 &&-0.606 \left( -0.61,  -0.61,  0.26,  0.26, 0.26, 0.26  \right),
\label{12-delet}
\end{eqnarray}
when the link between nodes $1$ and $2$ is deleted, and 
\begin{eqnarray}
|1\rangle =&&-0.707 \left(  -0.71,  0, 0,  0.71,  0, 0\right)\nonumber\\
                 &&-0.408 \left(  -0.41,  0,  0,  -0.41, 0.82, 0  \right)\nonumber\\
                 &&-0.289 \left(  -0.29,  0,  0,  -0.29, -0.29, 0.87  \right)\nonumber\\
                 &&+0.428 \left(    0.43,0.36,  0.36,  0.43, 0.43, 0.43  \right)\nonumber\\
                 &&+0.258 \left( 0.26,-0.61,-0.61 ,  0.26, 0.26, 0.26  \right)
\label{23-delet}
\end{eqnarray}
when the the link between nodes $2$ and $3$ is deleted. Finally,
\begin{eqnarray}
|1\rangle =&&-0.707 \left(  -0.71,  0, 0, 0, 0,  0.71\right)\nonumber\\
                 &&+0.365 \left(  0.36,  0.43,  0.43,  0.43, 0.43, 0.36  \right)\nonumber\\
                 &&0.606 \left( 0.61,  -0.26,  -0.26,  -0.26, -0.26, 0.61  \right)
\label{16-delet}
\end{eqnarray}
in the case of cutting the link between nodes $1$ and $6$. 

Equations (\ref{12-delet}) and (\ref{23-delet}) explicitly show the vanishing contribution of the node 
$6$ to some of the terms in the expansion, while Eq. (\ref{16-delet}) 
indicates that node $6$ contributes to all the components of the initial wave packet 
and in this case all the energy can be transmitted to the sink through this node.
We also checked that edge deletion between all the other nodes preserves localization in the resulting network. 
Figure \ref{edge-delet2} represents the block diagonalized structure of the stationary density matrices and the existence of localized states for some of these cases.

\begin{figure}[b]
\centering
\includegraphics[width=0.49\columnwidth]{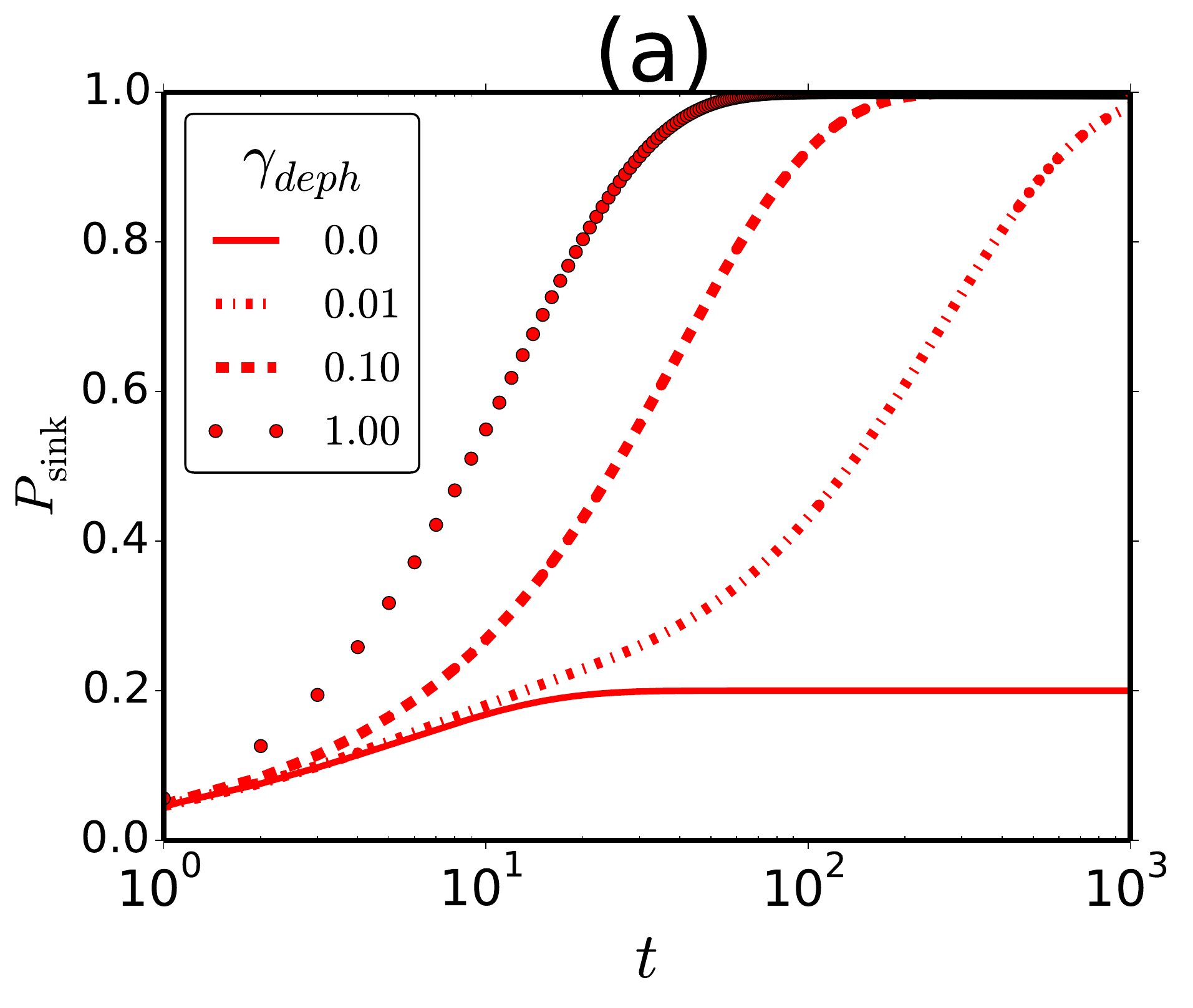}
\includegraphics[width=0.49\columnwidth]{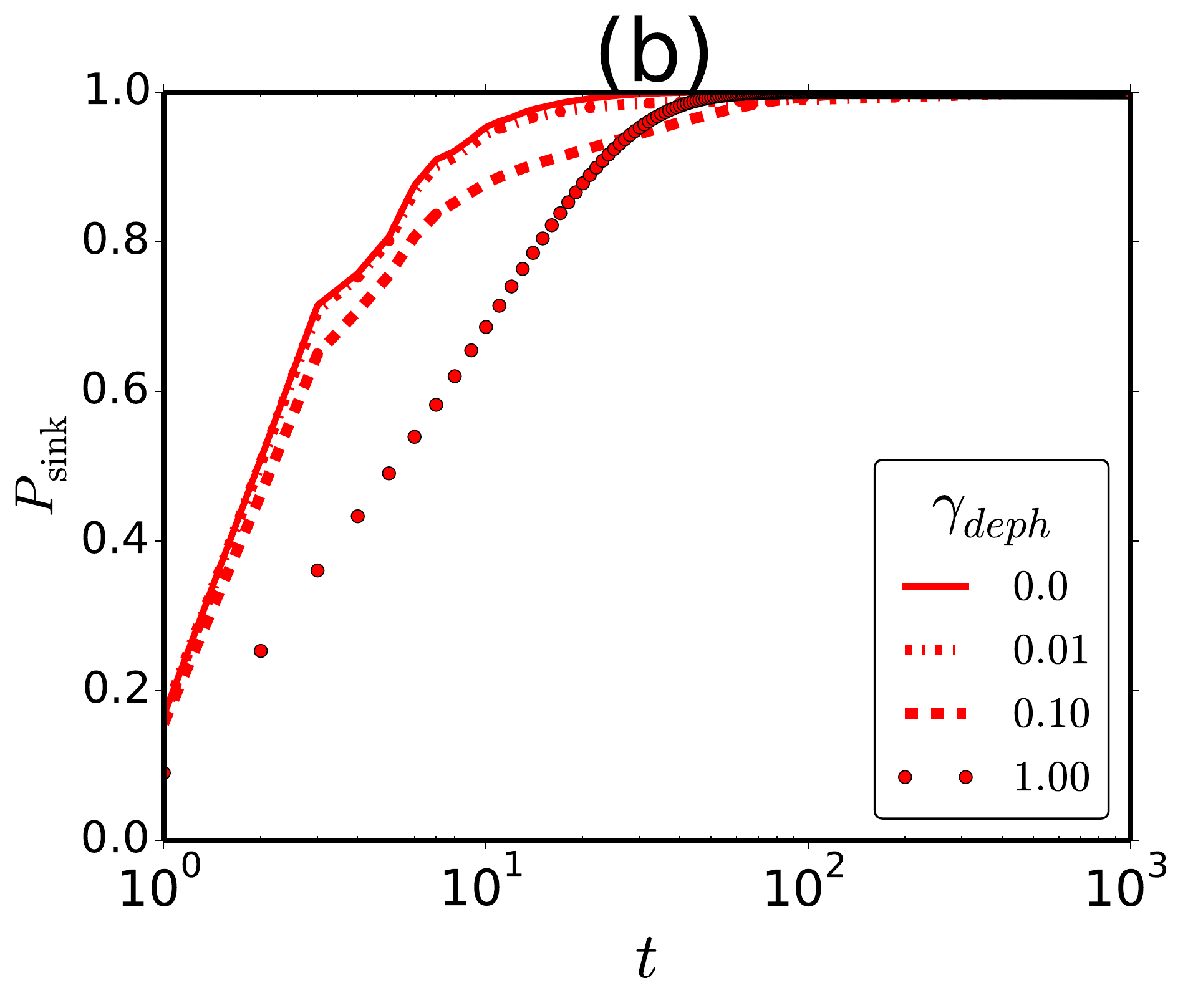}\\
\includegraphics[width=0.49\columnwidth]{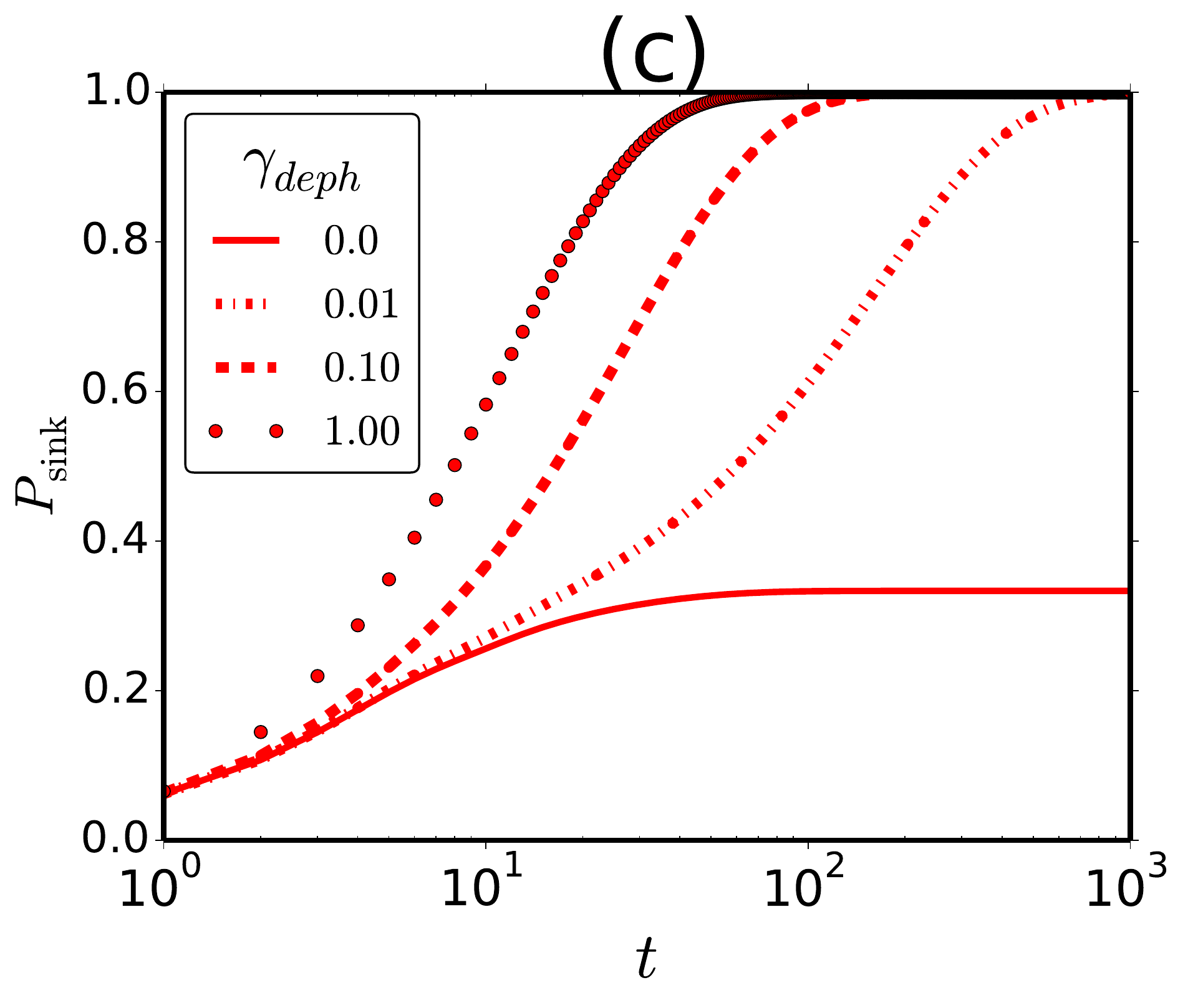}
\includegraphics[width=0.49\columnwidth]{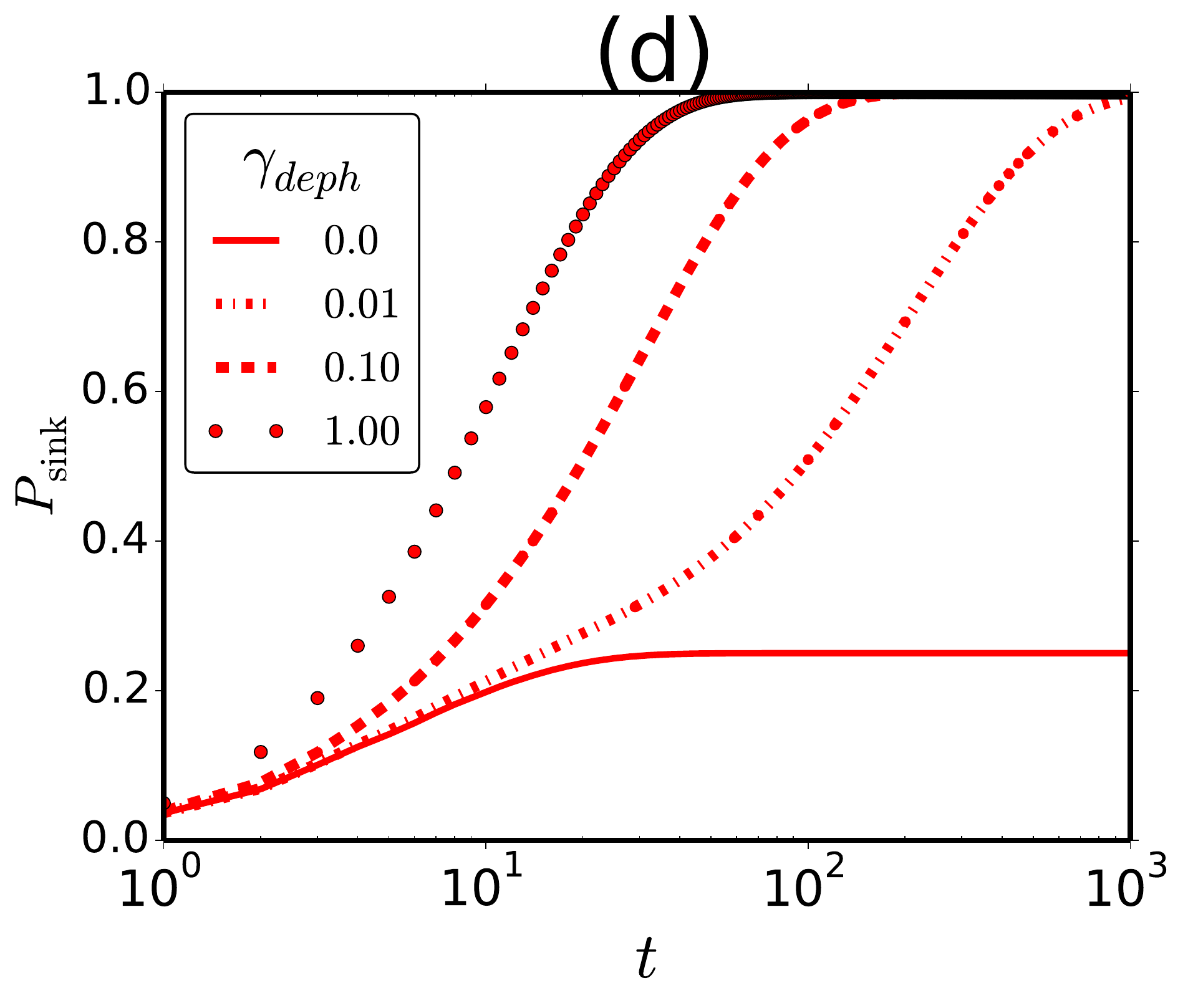}
    \caption{Sink population versus time (in logarithmic scale) in the absence (solid line) and presence of dephasing with different values of the 
dephasing coefficient $\gamma_{\rm deph}=0.01$ (dashed-dotted line), $0.1$ (dashed line) and $1.0$ (dotted line), for (a) a fully connected network, and networks with deletion of the edges (b)  $1-6$, (c) $2-4$ and (d) $3-6$. It can be seen that in the cases with localization, the influence of noise is constructive and reduces the saturation time, while for the case without localization (i.e. panel (b)) noise increases the saturation time. 
Thus noise has a constructive influence on the cases with localization.}
    \label{dephasing889}
\end{figure}


\section{Influence of dephasing}
\label{dephasing}

Next we investigate the influence of a noisy environment on quantum energy transport in networks. 
In most practical cases the networks are open ~\cite{Kossakowski,10} and interact with the environment,  which is expected to
reduce quantum coherence and constructive interference. Dephasing effects can be incorporated in the master equation in the framework of the Lindblad operators as follows~\cite{plenio.New.J.Phys.2008,hassan}:
\begin{equation}
 L_{\rm deph}\rho=\sum_{n=1}^{N}\gamma_{\rm deph}[2{\sigma}^+_n{\sigma}^-_n\rho{{\sigma}^+_n{\sigma}^-_n}-\big\{{{\sigma}^+_n{\sigma}^-_n,\rho}\big\}]
 \label{dephase}
 \end{equation}
where $\gamma_{\rm deph}$ is the dephasing rate coefficient.

Adding Eq. (\ref{dephase}) to the master equation (\ref{master}) and integrating this equation as before gives energy transfer in the presence of noise.  
Figure \ref{dephasing889} represents the time dependence of the sink population in the absence and presence of noise with different values of the dephacing rate coefficient. 
The results show that in cases where the wave packet is partially localized in the network, such as in the fully connected (Fig.~\ref{dephasing889}(a)), $2-4$ edge deleted (Fig.~\ref{dephasing889}(c)), and $3-6$ edge deleted (Fig.~\ref{dephasing889}(d)) networks,  the system efficiency monotonically increases (both in magnitude and speed) as dephasing is increased. Therefore, in these networks the 
noise has a constructive influence on energy transport by reducing quantum coherence.   
However, when the hopping between the initial node and the one connecting to the sink  ($1-6$) is blocked, 
the wave packet localization is already destroyed and in this case the noise slows down the speed of energy transfer to the sink as illustrated in 
Fig.~\ref{dephasing889}(b).



\section{Saturation Time}
\label{saturation}

\begin{figure}[t]
\includegraphics[width=0.8\columnwidth, height=0.7\columnwidth]{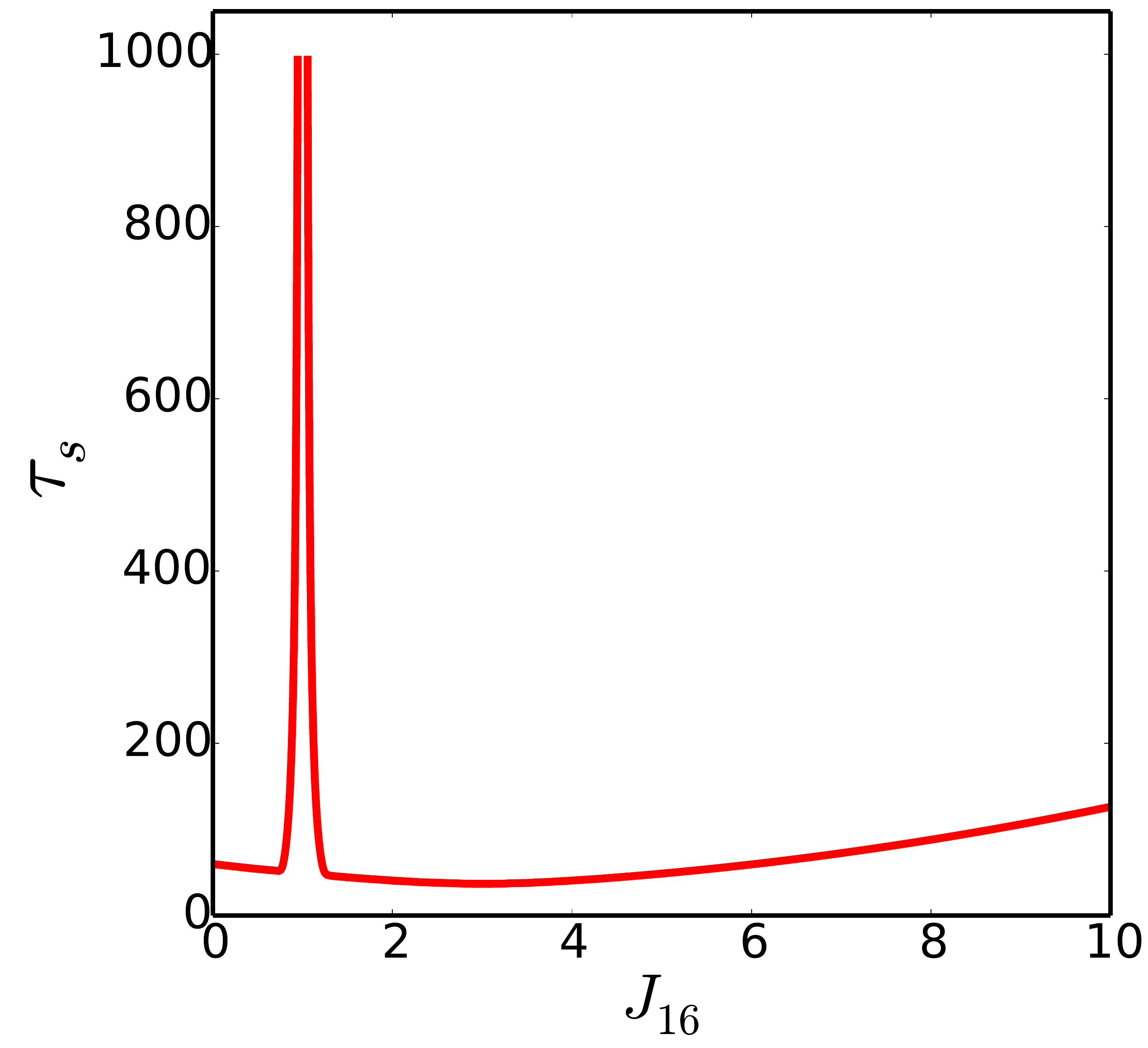}
    \caption{(Color online)  Dependence of the saturation time $\tau_s$ on the hopping rate $J_{16}$. The saturation time is minimized by $J^*_{16} \approx 3.04$. }
    \label{sat}
\end{figure}

The results presented here show that the link which connects the initial node to the final node plays a fundamental role in 
EET, and as Fig.~\ref{pop-J16} shows, when $J_{16}$ is 
changed from $0.0$ to $0.945$ or from $1.055$ to $2.0$, complete energy transfer is obtained. 
An obvious question then concerns finding a value of $J^*_{16}$ which is optimal in the sense 
that the {\it saturation time} $\tau_s$ is minimized. We define $\tau_s$ to be the time after which the injected energy is transferred to the sink in the absence of dephasing. 
In Fig. \ref{sat} we show numerical results for $\tau_s$ when varying the hopping rate  $J_{16}$. 
The nontrivial result here is that except for the peak caused by localization, 
the saturation time decreases with increasing $J_{16}$
attaining a minimum which is numerically determined to be $J^*_{16} \approx 3.04$. 

\begin{figure}[b]
\includegraphics[width=0.90\columnwidth]{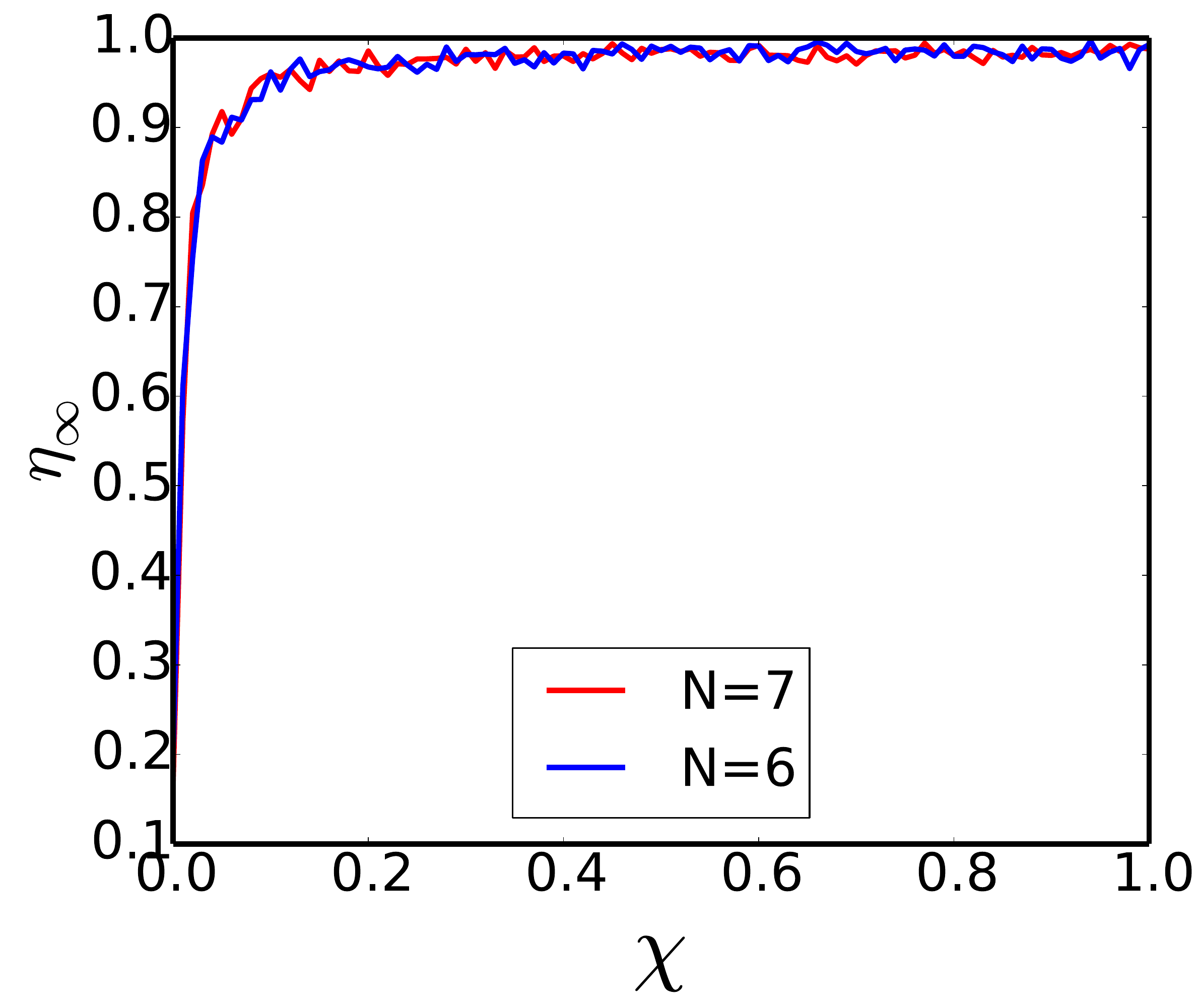}
\caption{System efficiency versus disorder strength}
\label{effikai}
\end{figure}

\section{Influence of Off-Diagonal Disorder}
\label{Off-DiagonalDisorder} 

In addition to edge deletion considered in the previous sections, it's also possible to introduce disorder
explicitly into the hopping integral parameters between the network nodes. In such networks it is expected that
the efficiency is strongly affected. We have studied this issue by introducing disorder
to the couplings by changing $J_{nm}$ to $J_{nm}(1 + \delta)$, where $\delta$ is a dimensionless random number with a uniform distribution in the interval $[-\chi, \chi]$. In Fig. 7 we show numerical
results for the efficiency parameter $\eta_{\infty}$ as a function of $\chi$ for two fully connected networks with $N=6$ and $N=7$ nodes. As expected, even a relatively small amount of disorder significantly increases
the efficiency in an FCN.

Finally, we have also studied the efficiency of edge-deleted networks in the case of $N=6$ nodes with
off-diagonal disorder and dissipation in the network, as given by the following term in the master equation:
\begin {equation} 
L_{\rm diss}\rho=\sum_{n=1}^{N}\gamma_n[2{\sigma}^-_n\rho{{\sigma}^+_n}-\big\{{{\sigma}^+_n{\sigma}^-_n,\rho}\big\}].
\label{dissipation}
\end {equation}
We have considered a large number of networks ($28$ in total), ranging from the case
where there are six connections between the nodes
up to the FCN, with 
fixed $\gamma_n=0.01$ for all sites, as a function of the disorder strength $\chi$ between $0 \le \chi \le 0.3$. 
While the detailed dependence of $\eta_\infty$ on $\chi$ is complicated and depends on the 
topology of the network, we find that for networks for which $\eta_{\infty}(\chi=0)$ is close to unity to start with,
increasing $\chi$ does not have much effect, as expected. In some cases, however, there's a slight approximately
linear {\it decrease} in the efficiency, up to about $10$ \% from $\chi=0$ to $\chi=0.3$. 
On the other hand, for networks where 
$\eta_{\infty}(\chi=0)$ is low (less than $0.4$), there's an approximately linear increase in the efficiency up to
about twice its value at zero disorder (e.g. in the case of the FCN). More detailed results of these studies will be published elsewhere.

\section{Summary and Conclusions}
\label{conclusion}

In summary, we have demonstrated  in this work that quantum transport of an energy excitation  in a symmetric network of fully connected two-state objects,  described by a time reversal symmetric Hamiltonian,
is highly inefficient. Whenever an excitation is injected into a node which is not directly connected to the sink,  a wave packet travelling through closed loop paths
arrives in-phase to the initial position due to the time reversal symmetry, which leads to destructive interference of the transition amplitudes at the initial site. 
The corresponding localization of the excitation energy wave packets inside the network is the reason for 
inefficiency of transport in a symmetric, fully connected network. 
Reducing the symmetry by eliminating hopping between nodes not directly connected to the sink preserves the localized states and hence does not significantly increase the system efficiency. 
We show that an efficient way to improve the efficiency of energy transfer is to introduce asymmetry between the initial node and the one directly linked to the sink. 
The rate of energy transfer can be minimized by a nontrivial value of the hopping rate $J_{16}$.
We have also
included dephasing to the network and shown, as expected, 
that the presence of a dephasing noise increases the system efficiency. The same effect can be obtained by introducing off-diagonal disorder in the
hopping integral parameters between the nodes.
Finally, we note that we have also carried out similar calculations for the cases $N=5$ and $N=15$ to verify the conclusions presented here. Our results should be useful in designing quantum networks for novel applications. 

{\bf Acknowledgements:}
T.A-N. has been supported in part by the Academy of Finland through its COMP CoE grant no. 251748. 
We wish to thank Leonardo Novo for useful comments.

\end {document}